\documentclass[fleqn,usenatbib]{mnras}

\usepackage{mathptmx}
\usepackage[T1]{fontenc}

\usepackage{graphicx}	% Including figure files
\usepackage{amsmath}	% Advanced maths commands
\usepackage{amssymb}	% Extra maths symbols

\title[The ultra-hot-Jupiter KELT-16\,b]{
The ultra-hot-Jupiter KELT-16\,b: Dynamical Evolution and Atmospheric Properties}

\author[L. Mancini et al.]{
\hspace{-0.2cm}
L.\ Mancini$^{1,2,3,4}$\thanks{E-mail:\href{lmancini@roma2.infn.it}{lmancini@roma2.infn.it}},
J.\ Southworth$^{5}$,
L.\ Naponiello$^{1,6}$,
\"{O}.\ Ba\c{s}t\"{u}rk$^{7}$,
D.\ Barbato$^{8,3}$,
F.\ Biagiotti$^{1}$,
I.\ Bruni$^{9}$,
\newauthor
L.\ Cabona$^{10}$,
G.\ D'Ago$^{11}$,
M.\ Damasso$^{3}$,
A.\ Erdem$^{12,13}$,
D.\ Evans$^{5}$,
Th.\ Henning$^{2}$,
O.\ \"{O}zt\"{u}rk$^{12,13}$,
D.\ Ricci$^{14}$,
\newauthor
A.\ Sozzetti$^{3}$,
J.\ Tregloan-Reed$^{15}$,
S. Yal\c{c}{\i}nkaya$^{7}$
\\
$^{1}$Department of Physics, University of Rome ``Tor Vergata'', Via della Ricerca Scientifica 1, I-00133, Rome, Italy\\
$^{2}$Max Planck Institute for Astronomy, K\"{o}nigstuhl 17, D-69117, Heidelberg, Germany \\
$^{3}$INAF -- Turin Astrophysical Observatory, via Osservatorio 20, I-10025, Pino Torinese, Italy \\
$^{4}$International Institute for Advanced Scientific Studies (IIASS), Via G.\ Pellegrino 19, I-84019, Vietri sul Mare (SA), Italy \\
$^{5}$Astrophysics Group, Keele University, Keele ST5 5BG, UK \\
$^{6}$Department of Physics and Astronomy, University of Florence, Via Sansone 1, I-50019, Sesto Fiorentino (FI), Italy \\
$^{7}$Ankara University, Faculty of Science, Astronomy \& Space Sciences Department, Tandogan, TR-06100, Ankara, Turkey \\
$^{8}$Observatoire de Gen\`{e}ve, Universit\`{e} de Gen\`{e}ve, 51 Chemin Pegasi, 1290 Sauverny, Switzerland \\
$^{9}$INAF -- OAS, Osservatorio di Astrofisica e Scienza dello Spazio di Bologna, Via P. Gobetti 93/3, I-40129, Bologna, Italy \\
$^{10}$INAF -- Brera Astronomical Observatory, Via E. Bianchi 46, I-23807, Merate (LC), Italy \\
$^{11}$Instituto de Astrof\'{i}sica, Pontificia Universidad Cat\'{o}lica de Chile, Av. Vicu\~{n}a Mackenna 4860, 7820436 Macul, Santiago, Chile \\
$^{12}$Department of Physics, Faculty of Arts and Sciences, \c{C}anakkale Onsekiz Mart University, Terzio\u{g}lu Kamp\"{u}s\"{u}, TR-17020, \c{C}anakkale, Turkey \\
$^{13}$Astrophysics Research Centre and Observatory, \c{C}anakkale Onsekiz Mart University, Terzio\u{g}lu Kamp\"{u}s\"{u}, TR-17020, \c{C}anakkale, Turkey \\
$^{14}$INAF -- Padova Astronomical Observatory, Vicolo dell'Osservatorio 5, I-35122, Padova, Italy \\
$^{15}$Instituto de Investigaci\'{o}n en Astronomia y Ciencias Planetarias, Universidad de Atacama, Campus Paulino del Barrio Avenida Copayapu 485, Copiap\'{o}, Chile
}

\date{Accepted XXX. Received YYY; in original form ZZZ}

\pubyear{2021}

\usepackage{newtxtext,newtxmath}

\begin{document}
\label{firstpage}
\pagerange{\pageref{firstpage}--\pageref{lastpage}}
\maketitle

\begin{abstract}
We present broad-band photometry of 30 planetary transits of the ultra-hot Jupiter KELT-16\,b, using five medium-class telescopes. The transits were monitored through standard $B,\,V,\,R,\,I$ filters and four were simultaneously observed from different places, for a total of 36 new light curves. We used these new photometric data and those from the TESS space telescope to review the main physical properties of the KELT-16 planetary system. Our results agree with previous measurements but are more precise. We estimated the mid-transit times for each of these transits and combined them with others from the literature to obtain 69 epochs, with a time baseline extending over more than four years, and searched for transit time variations. We found no evidence for a period change, suggesting a lower limit for orbital decay at $8$\,Myr,
with a lower limit on the reduced tidal quality factor of $Q^{\prime}_{\star}>(1.9 \pm 0.8) \times 10^5$ with $95\%$ confidence.
We built up an observational, low-resolution transmission spectrum of the planet, finding evidence of the presence of optical absorbers, although with a low significance. Using TESS data, we reconstructed the phase curve finding that KELT-16\,b has a phase offset of $ 25.25 \pm 14.03$\,$^{\circ}$E, a day- and night-side brightness temperature of $3190 \pm 61$\,K and $2668 \pm 56$\,K, respectively. Finally, we compared the flux ratio of the planet over its star at the TESS and Spitzer wavelengths with theoretical emission spectra, finding evidence of a temperature inversion in the planet's atmosphere, the chemical composition of which is preferably oxygen-rich rather than carbon-rich.
\end{abstract}


%%%%%%%%%%%%%%%%%%%%%%%%%%%%%%%%%%%%%%%%%%%%%%%%%%

\section{Introduction}
\label{sec:intro}
Although hot Jupiters were the first exoplanets that have been discovered more than twenty years ago \citep{mayor:1995}, there are still many open questions related to this class of planets. It is not well established, for example, the predominant physical mechanism that caused these gas planets to form, as well as how did they come to migrate and occupy such short-period orbits ($\sim 1$\,day) around their host stars\footnote{The two leading theories, which can explain the shrinking of the orbit of a giant planet, are ($i$) early interactions between a giant planet and the protoplanetary disk of its parent star and ($ii$) planet-planet scattering.}. The future evolution of a hot Jupiter is also unclear and difficult to predict exactly but is surely linked to that of its star and its interaction with it. This because stellar tidal forces become particularly relevant when a giant planet is at a distance of roughly $0.1$\,au from its host star. 

The gravitational interaction forces a hot Jupiter to change its rotation rate and circularise its orbit as a result of energy exchange and heat dissipation. This can lead to a spin-orbit synchronization and eventually to orbital decay 
(e.g., \citealt{hut:1980,rasioford:1996,sasselov:2003,ferraz:2008,levrard:2009}). When the planet approaches the Roche limit, it begins to lose mass, in favour of its star, via a mechanism known as \emph{Roche-lobe overflow}. At the Roche limit, tidal forces overcome the gravity of the planet itself, which may then be partially or totally disrupted~\citep{dosopoulou:2017}. In case of planetary systems, the Roche-limit separation can be written as $a_{\rm R}\approx 2.165\,R_{\rm p}(M_{\star}/M_{\rm p})^{1/3}$ \citep{paczynski:1971,matsumura:2010}.

The frictional processes in the star lead to the dissipation of orbital energy, whose efficiency can be parameterised by the \emph{tidal dissipation quality factor}, $Q_{\star}$, which can be expressed as the ratio between the energy in the orbit of a binary system and the amount of energy lost in each orbit due to internal friction~\citep{goldreich:1966,matsumura:2010,barnes:2011,wilkins:2017}. In the following, we will use the \emph{modified tidal dissipation quality factor}, $Q^{\prime}_{\star}$, defined by \citet{ogilvie:2007} as $Q^{\prime}_{\star} \equiv \, 1.5 \,Q_{\star}/k_{2}$, where $Q_{\star}$ is the stellar tidal quality factor and $k_{2}$ is the Love number for the second-order harmonic potential~\citep{love:1944}, which is related to the star's density profile.

The rate of the energy dissipation is related to the physical and orbital parameters of the planet as well as the spectral class, metallicity and rotational evolution of the parent star \citep{barker:2010,ogilvie:2014,essick:2016,gallet:2017,bolmont:2017}. Therefore, the orbital-decay time of a hot Jupiter is very difficult to estimate, but it should occur on a very long timescale \citep{rasio:1996,lai:2012}, with evidence that the tidal destruction of hot Jupiters happens during the main-sequence lifetimes of their host stars \citep{chernov:2017,hamer:2019}.

Nevertheless, it is useful to try to determine the value of $Q^{\prime}_{\star}$, since it would provide insights into the interior parameters of the star and be an empirical test to current dynamical models of close-in planets and tidal stability. Theoretical studies of tidal evolution of hot Jupiters, based on different initial conditions and assumptions, suggest $10^7<Q^{\prime}_{\star}<10^{10}$ \citep{ogilvie:2014}, $10^6<Q^{\prime}_{\star}<10^9$ \citep{penev:2011,penev:2012} or $Q^{\prime}_{\star}\lesssim 10^7$ \citep{hamer:2019}. 
A statistical inference based on a large sample of hot Jupiters yielded $Q^{\prime}_{\star}\sim 10^8$ \citep{collier:2018}.
Instead, \citet{jackson:2008} found that the distribution of initial eccentricities of close-in planet ($a < 0.2$\,au) matches that of the general population, the best agreement being with $Q^{\prime}_{\star} \simeq 10^{5.5}$. A similar value of the tidal dissipation quality factor ($Q^{\prime}_{\star}\simeq 10^5-10^6$) was estimated by \citet{essick:2016} for systems with solar-type host star, planet mass $M_{\rm p} \gtrsim 0.5 \, M_{\rm Jup}$ and orbital period $P \lesssim 2$\,days. Therefore, such a planet can decay on smaller timescales than the main-sequence lifetime of its host. 

%----------------------------------------
\subsection{Transit timing variations for estimating $Q^{\prime}_{\star}$ }
%\subsection{TTTVs for estimating modified tidal dissipation quality factor }
\label{sec:intro_2} %
%----------------------------------------

Under several assumptions and using Kepler's third law \citep{goldreich:1966,levrard:2009,matsumura:2010,birkby:2014,collier:2018}, one can deduce that the tidal dissipation quality factor is related to the time derivative $\dot{P}$ of the planet's orbital period through
\begin{equation}
\label{eq:quality-factor}
Q^{\prime}_{\star} = -\frac{27\piup}{2 \dot{P}}
\frac{M_{\rm {p}}}{M_{\star}} \left(\frac{R_{\star}}{a}\right)^{5},
\end{equation}
where $M_{\rm p}$ is the mass of the orbiting planet, $a$ is the semi-major axis of the orbit, while $M_{\star}$ and $R_{\star}$ are the mass and the radius of the parent star, respectively. 

Variations of the orbital period can be found for those hot Jupiters that transit in front of their parent stars. The methodology is to systematically measure their mid-transit times and search for the so-called transit timing variations (TTVs). Such measurements are not difficult to achieve, although special attention is required for certain aspects, such as the clocks connected with the telescopes, the fit of the light curves, combining measurements made with different instruments, etc.

After excluding all possible sources of uncertainty, possible detection of TTVs in a transiting planetary system can be explained by several scenarios like the presence of additional bodies in the system, an apsidal precession or an unstable orbit, the latter resulting from tidal forces generated by the parent star. The orbital decay results in a non-zero time derivative of the orbital period, which can be measured, for example, by fitting transit timings to a quadratic ephemeris or by detecting long-term deviations from the linear ephemeris.

Constraining $Q^{\prime}_{\star}$ is, therefore, possible for a single transiting planet, but not easy to achieve, as it usually requires systematic observations of planetary-transit events over many years and high-precision photometry. Indeed, even though many hot Jupiters have been extensively monitored during the last two decades, only a few and controversial detections of orbital decay have been claimed to date (e.g., \citealt{adams:2010,murgas:2014,blecic:2014,hoyer:2016,jiang:2016,pablo:2017,southworth:2019}). Instead, \citet{maciejewski:2018b} found no departure from a constant-period model for WASP-18\,b, despite this being one of the best candidates.

The most convincing case for the detection of orbital decay is that of the hot Jupiter WASP-12\,b \citep{hebb:2009}, whose transits were systematically monitored almost immediately after the announcement of its discovery, straightway highlighting evidence of a decreasing orbital period \citep{maciejewski:2013,maciejewski:2016}, which was later confirmed \citep{patra:2017,maciejewski:2018} and recently ascribed to the orbital decay of the planet \citep{yee:2020,turner:2021}. In particular, \citet{turner:2021} estimated for WASP-12 that $Q^{\prime}_{\star} = 1.39 \pm 0.15 \times 10^5$, a low value if compared with most of the theoretical predictions, but in agreement with the previous estimate, $Q^{\prime}_{\star}= 1.6 \pm 0.2 \times 10^5$, from \citet{patra:2020}. %

\begin{figure*}
\includegraphics[width=16cm]{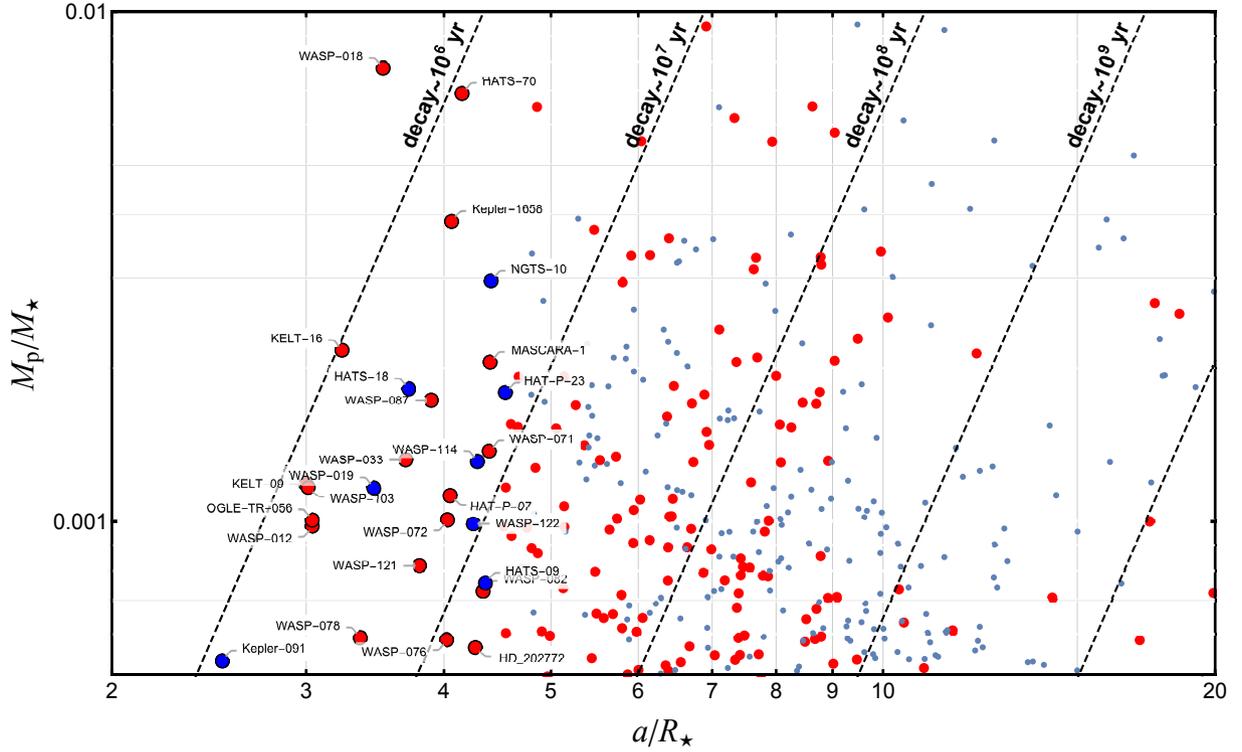}
\caption{$M_{\rm p}/M_{\star}$ versus $a/R_{\star}$ diagram for the known transiting planetary systems (the values of the parameters were taken from TEPCat). The points representing the systems are marked with two different colours, according to the temperature of the host stars, i.e. red for those with $T_{\rm eff} > 6000$\,K and blue for the others. Error bars have been suppressed for clarity. Dark lines show where the decay timescale is constant and have been calculated using Eq.~(\ref{eq:quality-factor}), considering a nominal value of $Q^{\prime}_{\star}=10^6$. The first twenty most favourable targets are highlighted and labelled. Figure inspired by \citet{collier:2018} and \citet{patra:2020}.
}
\label{fig:diagram_1}
\end{figure*}
%
%----------------------------------------
\subsection{Planets that may incur in orbital decay}
\label{sec:intro_3} %
%----------------------------------------

\begin{figure}
\includegraphics[width=\hsize]{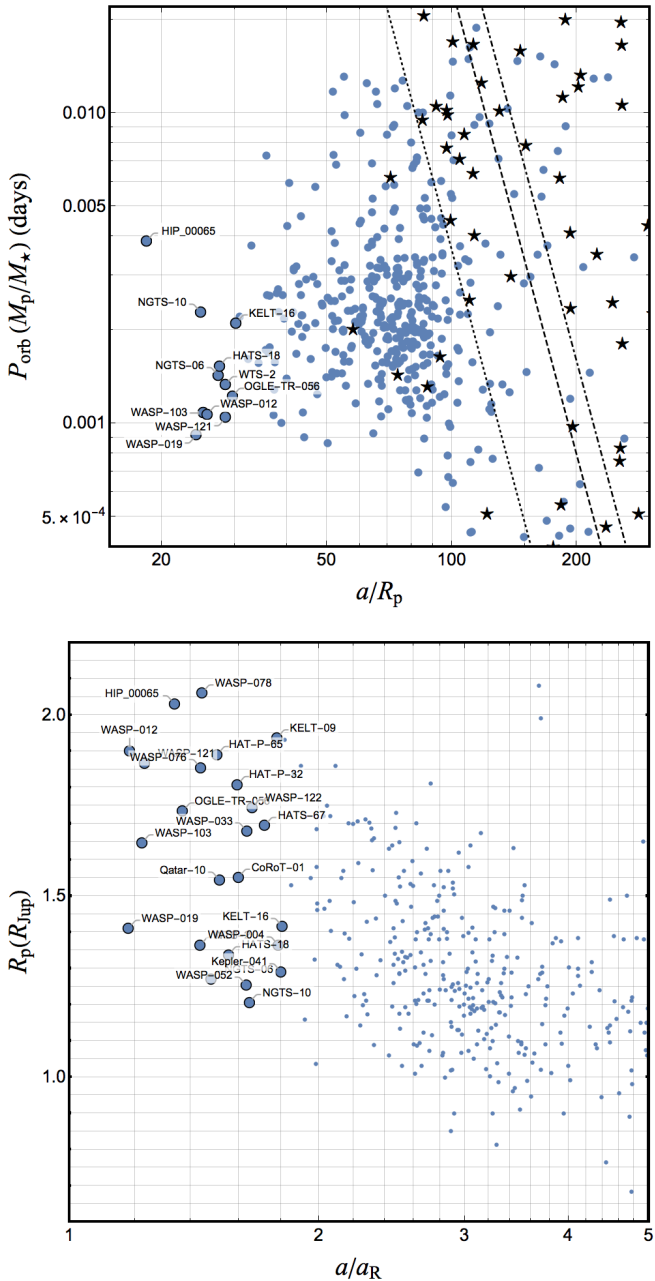}
\caption{{\it Top panel:} Modified tidal diagram of known transiting exoplanets. Circles display the position of planets with eccentricity $e < 0.1$, while five-pointed stars those with $e > 0.1$. Error bars have been suppressed for clarity. The dotted, dashed and dash-dotted lines display the 1, 7, and 14\,Gyr circularisation timescales for $Q^{\prime}_{\rm p}=10^6$ and $e=0$, respectively. Figure inspired from \citet{bonomo:2017}. {\it Bottom panel:} Distribution of the known larger transiting hot Jupiters versus the orbital semi-major axis in units of Roche radii $a_{\rm R}$. The values of the parameters were taken from TEPCat. The planets close to tidal disruption are highlighted and labelled.}
\label{fig:diagram_2}
\end{figure}
Considering Eq.~(\ref{eq:quality-factor}), it is possible to appreciate how much WASP-12\,b is a very good candidates for detecting orbital decay, as we can also see from Fig.~\ref{fig:diagram_1}, where we plotted $M_{\rm p}/M_{\star}$ versus $a/R_{\star}$ for the known transiting planets\footnote{Values taken from the Transiting Extrasolar Planet Catalogue (TEPCat), which is available at {\tt http://www.astro.keele.ac.uk/jkt/tepcat/} \citep{southworth:2011}.}. Another worthwhile plot to understand the impact of tidal interactions on planetary orbits is the \emph{tidal diagram} \citep{pont:2011}. The circularisation time $\tau_{\rm e}$ can be written as \citep{matsumura:2008,bonomo:2017}\footnote{Note that the $\pi$ in Eq. (1) of \citealt{bonomo:2017} was incorrectly reversed, as confirmed by Bonomo (2021), private communication.}
\begin{equation}
\label{eq:tidal-diagram}
\tau_{\rm e}=P_{\rm orb}\frac{M_{\rm p}}{M_{\star}} \frac{2 Q^{\prime}_{\rm p}}{63 \pi}\left(\frac{a}{R_{\rm p}}\right)^{5},
\end{equation}
where $R_{\rm p}$ is the radius of the planet and $Q^{\prime}_{\rm p}$ is the planetary modified tidal quality factor, which is defined as $Q^{\prime}_{\rm p}=3Q_{\rm p}/2k_2$, $Q_{\rm p}$ being the planet tidal quality factor. Following \citet{bonomo:2017}, we considered a \emph{modified tidal diagram} by plotting $P_{\rm orb}(M_{\rm p}/M_{\star})$ versus $(a/R_{\rm p})$ in the top panel of Fig.~\ref{fig:diagram_2}, so that the circularisation isochrones do not depend on the planetary orbital periods. From this plot, it is possible to appreciate how most of the eccentric planets, with $e>0.1$ (black stars), are found beyond the 1\,Gyr circularisation isochrone. Finally, in the bottom panel of Fig.~\ref{fig:diagram_2} we show the transiting hot Jupiters with the smallest ratio $a/a_{\rm R}$.

Examining Figs.~\ref{fig:diagram_1} and \ref{fig:diagram_2}, KELT-16\,b is one of the most promising hot Jupiters whose transits are worth to precisely record for searching possible tidal-decay signatures. This planet is the object we selected for the study that we present in this work and its properties are described in Sect.\,2. The paper is organized as follows.
In Sect.\,3 we present new photometric follow-up observations of KELT-16\,b transits. The data reduction is also described in Sect.\,3. The analysis of the light curves is presented in Sect.\,4, while, in Sect.\,5, we analysed the transit times and investigated the possibility of a decay of the planetary orbit. In Sect.\,6 we revise the main physical properties of the KELT-16 planetary system and investigated the atmosphere composition of its planet. Finally, we summarise our results in Sect.\,7.

%----------------------------------------
\section{The case study: KELT-16 b}
\label{sec:intro_kelt16} %
%----------------------------------------
KELT-16\,b was discovered by \citet{oberst:2017}. It has $M_{\rm p}=2.71\,M_{\rm Jup}$, $R_{\rm p}=1.38\,R_{\rm Jup}$ and orbits in $0.97$~days at $\approx 1.8\,a/a_{\rm R}$ around the bright ($V=11.7$\,mag) F7\,V star KELT-16 (aka TYC\,2688-1839-1, aka TOI-1282) in the thin disk of the Galaxy, on the outskirts of the Cygnus Loop nebula, a supernova remnant. This star has a widely separated bound companion (a $V=19.6$\,mag, M3\,V type red dwarf), with a  minimum separation of $\approx 300$\,au and an effective temperature of $T_{\rm eff}\approx3400$\,K \citep{oberst:2017}. This suggests the possibility to have Kozai-Lidov (KL) oscillations, which could be responsible for having driven KELT-16\,b to its current orbit.

With an equilibrium temperature of $\approx 2450$\,K, KELT-16\,b can be considered as an ultra-hot Jupiter. By using an orbital evolution model, \citet{oberst:2017} showed that KELT-16\,b could be tidally disrupted in roughly $5.5 \times 10^5$\,yr, if its parent stars has a value of $Q^{\prime}_{\star}\sim 10^5$, i.e. similar to that of WASP-12, which was mentioned above (see Sect.~\ref{sec:intro}). 

Due to its particular orbital period ($\approx 1$\,day), there are several blocks of consecutive nights in which KELT-16\,b undergoes transit events during a year. This is a rare and peculiar characteristic for a transiting planet, because it allows intensive and continuous monitoring of planetary-transit times for all these nights with ground-based telescopes, also simplifying the schedule of observations.

\begin{figure}
\includegraphics[width=\hsize]{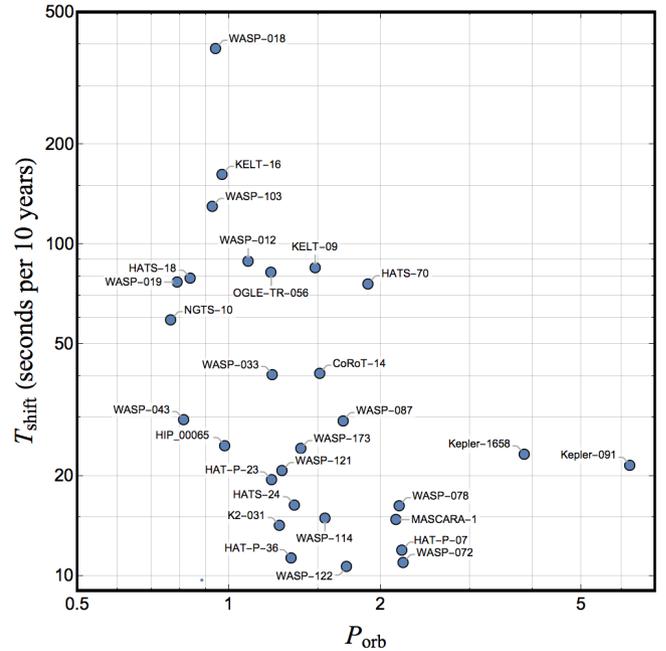}
\caption{$T_{\rm shift}-P_{\rm orb}$ diagram for the known transiting hot Jupiters, whose transits should present the largest mid-transit time shifts after $10$\,yr. These planets are labelled. Data taken from TEPCat. Figure inspired from \citet{maciejewski:2018}.}
\label{fig:diagram_3}
\end{figure}

\citet{maciejewski:2018} selected this target as one of the best candidates in the northern hemisphere for detecting planet-star tidal interactions. Their selection included planets for which the $T_{\rm shift}$, the predicted cumulative shift in transit times, was greater than $30$\,s after 10 years. Figure~\ref{fig:diagram_3} shows the $T_{\rm shift}-P_{\rm orb}$ diagram for transiting hot Jupiters, where $T_{\rm shift}$ was derived from Eq.~(\ref{eq:quality-factor}) (see also \citealt{maciejewski:2018}), in which the position of KELT-16\,b, as well as those of other hot Jupiters, is highlighted.

Eleven transit observations were reported by \citet{maciejewski:2018}, who used an array of five different telescopes, with apertures between 0.6\,m and 2\,m. The corresponding mid-transit times were joined with those from the discovery paper \citep{oberst:2017}, finding that they were well fitted by a linear ephemeris. The same result was obtained by \citet{patra:2020}, who observed another two transits of KELT-16\,b with a 1.2\,m telescope.

Finally, a Spitzer/IRAC $4.5\,\mu$m phase curve was reported by \citet{bell:2021}, who estimated the day- and night-side temperature of the planet. 

%----------------------------------------
%\subsection{Contents}
%\label{sec:contents} %
%----------------------------------------
%The paper is organized as follows. 
%In Sect.\,2 we present new photometric follow-up observations of KELT-16\,b transits. The data reduction is also described in %Sect.\,2. The analysis of the light curves is presented in Sect.\,3, while, in Sect.\,4, we analysed the transit times and %investigated the possibility of a decay of the planetary orbit. In Sect.\,5 we revise the main physical properties of the %KELT-16 planetary system and investigated the atmosphere composition of its planet. Finally, we summarise our results in Sect.
%\,6.

%----------------------------------------
\section{Observations and data reduction}
\label{sec:observations} %
%----------------------------------------

%%%%%%%%%%%%%%%%%%%%%%%%%%%%%%%%%%%%%%%
% Table 1
%%%%%%%%%%%%%%%%%%%%%%%%%%%%%%%%%%%%%%%
\begin{table*}
%{\tiny
\centering
\caption{Details of the transit observations presented in this work. $N_{\rm obs}$ is the number of observations,
$T_{\rm exp}$ is the exposure time, $T_{\rm obs}$ is the observational cadence, and `Moon illum.' is the geocentric
fractional illumination of the Moon at midnight (UT). The aperture sizes are the radii of the software apertures
for the star, inner sky and outer sky, respectively. Scatter is the \emph{rms} scatter of the data versus a fitted
model. The last column specifies if the transit was simultaneously observed by more than one telescope.}
\label{tab:obs}
\resizebox{\hsize}{!}{
\begin{tabular}{lcccrcclcrccc} \hline
Telescope & Date of   & Start time & End time  &$N_{\rm obs}$ & $T_{\rm exp}$ & $T_{\rm obs}$ & ~~Filter & Airmass & Moon & Aperture & Scatter & Simultaneous  \\
               & first obs &    (UT)    &   (UT)    &              & (s)           & (s)           &        &         &illum.& radii (px) & (mmag)  \\
\hline
CA\,1.23\,m &      2017/06/19 & 23:58 & 03:46 &  191~ &  60 & 72 & Cousins $I$ & $1.31 \rightarrow 1.00 \rightarrow 1.01$  &  22\%  & 16,\,25,\,49  & 1.27 & no \\
CA\,1.23\,m &      2017/06/20 & 21:59 & 03:50 &  295~ &  60 & 72 & Cousins $I$ & $2.21 \rightarrow 1.00 \rightarrow 1.01$  &  14\%  & 17,\,26,\,48  & 1.67 & no \\
Cassini\,1.52\,m & 2017/06/21 & 21:56 & 02:41 &  235~ &  40/95 & 58/113 & Johnson $R$ & $1.33 \rightarrow 1.02 \rightarrow 1.06$  &  7\%  & 11,\,15,\,23  & 1.60 & yes \\
CA\,1.23\,m &      2017/06/21 & 22:12 & 04:11 &  230~ &  60/95 & 72/107 & Cousins $R$ & $1.95 \rightarrow 1.00 \rightarrow 1.04$  &  7\%  & 18,\,27,\,49  & 1.21 & yes \\
OARPAF\,80\,cm &   2017/06/21 & 22:18 & 02:23 &  225~ &  50 & 62 & Johnson $R$ & $1.03 \rightarrow 1.29$  &  7\%  & 14,\,15,\,24  & 2.15 & yes \\
CA\,1.23\,m &      2017/06/22 & 21:42 & 03:00 &  232~ &  55/95 & 67/107 & Cousins $R$ & $2.15 \rightarrow 1.00$  &  2\%  & 18,\,26,\,49  & 1.60 & yes \\
Cassini\,1.52\,m & 2017/06/22 & 22:18 & 02:36 &  230~ &  40/75 & 58/93 & Johnson $R$ & $1.36 \rightarrow 1.02 \rightarrow 1.07$  &  2\%  & 12,\,16,\,25  & 1.25 & yes \\
Cassini\,1.52\,m & 2017/06/23 & 22:00 & 01:15 &  134~ &  40/90 & 58/108 & Johnson $V$ & $1.40 \rightarrow 1.02$ & 0\% & 11,\,14,\,25  & 2.00 & no \\
T100\,1\,m &       2017/06/25 & 19:37 & 01:16 &  126~ &  110 & 155 & Bessel $R$ & $2.17 \rightarrow 1.01$  &  5\%  & 15,\,20,\,25  & 1.82 & no \\
SPM\,84\,cm &      2017/07/10 & 06:43 & 12:01 &  108~ &  150 & 162 & Bessel $R$ & $1.22 \rightarrow 1.00 \rightarrow 1.18$  &  95\%  & 20,\,22,\,27 & 2.40 & no \\
SPM\,84\,cm &      2017/07/11 & 05:57 & 11:58 &  135~ &  150 & 162 & Bessel $R$ & $1.39 \rightarrow 1.00 \rightarrow 1.18$  &  89\%  & 20,\,22,\,27 & 1.35 & no \\
Cassini\,1.52\,m & 2017/07/22 & 21:56 & 02:41 &  269~ &  40/95 & 58/113 & Johnson $V$ & $1.33 \rightarrow 1.02 \rightarrow 1.06$  &  0\%  & 11,\,15,\,23  & 1.60 & no \\
Cassini\,1.52\,m & 2017/07/24 & 20:51 & 02:07 &  113~ &  120/135 & 138/153 & Johnson $B$ & $1.26 \rightarrow 1.02 \rightarrow 1.24$  &  4\%  & 15,\,20,\,50  & 1.16 & no \\
Cassini\,1.52\,m & 2017/07/25 & 20:11 & 01:36 &  163~ &  95/120 & 113/138 & Johnson $B$ & $1.32 \rightarrow 1.02 \rightarrow 1.36$  &  9\%  & 13,\,20,\,52  & 1.00 & no \\
T100\,1\,m &       2017/07/27 & 18:17 & 22:19 &   92~ &  110 & 155 & Bessel $I$ & $1.67 \rightarrow 1.01$  &  24\%  & 30,\,40,\,45  & 1.44 & no \\
CA\,1.23\,m &      2017/08/20 & 22:52 & 04:38 &  287~ &  60 & 72 & Cousins $I$ & $1.05 \rightarrow 1.00 \rightarrow 1.68$  &  1\%  & 18,\,28,\,47  & 1.27 & no \\
CA\,1.23\,m &      2017/08/22 & 21:19 & 04:23 &  351~ &  60 & 72 & Cousins $R$ & $1.11 \rightarrow 1.00 \rightarrow 2.23$  &  2\%  & 18,\,30,\,50  & 1.45 & no \\
CA\,1.23\,m &      2017/08/23 & 21:02 & 02:53 &  295~ &  60 & 72 & Johnson $V$ & $1.35 \rightarrow 1.00 \rightarrow 1.93$  &  6\%  & 18,\,28,\,47  & 1.53 & no \\
CA\,1.23\,m &      2017/09/24 & 20:21 & 02:33 &  192~ &  100 & 112 & Johnson $V$ & $1.01 \rightarrow 1.00 \rightarrow 2.58$  &  23\%  & 15,\,24,\,50  & 1.34 & no \\
CA\,1.23\,m &      2017/09/25 & 19:15 & 02:00 &  174~ &  100/180 & 112/192 & Johnson $B$ & $1.07 \rightarrow 1.00 \rightarrow 2.12$  &  31\%  & 22,\,32,\,55  & 1.55 & no \\
CA\,1.23\,m &      2017/09/26 & 19:28 & 01:22 &  284~ &  50/100 & 62/112 & Cousins $R$ & $1.04 \rightarrow 1.00 \rightarrow 1.76$  &  40\%  & 16,\,25,\,50  & 1.87 & no \\
CA\,1.23\,m &      2017/09/29 & 18:00 & 04:35 &  226~ &  55/100 & 67/112 & Cousins $R$ & $1.01 \rightarrow 1.00 \rightarrow 2.42$  &  68\%  & 16,\,26,\,50  & 1.11 & no \\
CA\,1.23\,m &      2018/07/01 & 22:19 & 02:16 &  136~ &  90/120 & 102/132 & Cousins $I$ & $1.80 \rightarrow 1.00 \rightarrow 1.03$  &  87\%  & 12,\,35,\,55  & 1.06 & no \\
CA\,1.23\,m &      2018/07/02 & 21:43 & 03:17 &  139~ &  120 & 132 & Cousins $R$ & $1.81 \rightarrow 1.00 \rightarrow 1.02$  &  80\%  & 18,\,27,\,47  & 1.19 & yes \\
Cassini\,1.52\,m & 2018/07/02 & 21:23 & 02:42 &  122~ &  120 & 138 & Johnson $R$ & $1.32 \rightarrow 1.02 \rightarrow 1.12$  &  81\%  & 14,\,18,\,28  & 1.33 & yes \\
Cassini\,1.52\,m & 2018/07/03 & 21:20 & 23:56 &   70~ &  120 & 138 & Johnson $V$ & $1.31 \rightarrow 1.03$  &  72\%  & 12,\,18,\,25  & 2.10 & yes \\
CA\,1.23\,m &      2018/07/03 & 21:39 & 02:25 &  121~ &  120/130 & 132/142 & Cousins $R$ & $1.87 \rightarrow 1.00$  &  72\%  & 22,\,30,\,47  & 1.22 & yes \\
T100\,1\,m &       2019/07/12 & 22:03 & 01:20 &   79~ &  90 & 135 & Bessel $R$ & $1.06 \rightarrow 1.00 \rightarrow 1.07$  &  85\%  & 20,\,25,\,30  & 1.79 & no \\
T100\,1\,m &       2019/08/14 & 20:03 & 00:33 &  303~ &  30 & 45 & Bessel $R$ & $1.05 \rightarrow 1.00 \rightarrow 1.28$  &  96\%  & 13,\,21,\,26  & 7.13 & no \\
T100\,1\,m &       2019/09/16 & 18:50 & 23:07 &  297~ &  30 & 45 & Bessel $R$ & $1.01 \rightarrow 1.00 \rightarrow 1.48$  &  93\%  & 10,\,17,\,22  & 3.78 & no \\
T100\,1\,m &       2019/09/17 & 17:16 & 23:05 &  422~ &  30 & 45 & Bessel $R$ & $1.10 \rightarrow 1.00 \rightarrow 1.49$  &  87\%  & 12,\,17,\,22  & 2.88 & no \\
T100\,1\,m &       2020/05/20 & 21:26 & 01:42 &   85~ &  110 & 155 & Bessel $R$ & $2.55 \rightarrow 1.05$  &   3\%  & 17,\,33,\,46  & 1.73 & no \\
Cassini\,1.52\,m & 2020/07/19 & 20:26 & 02:28 &  263~ &  120 & 138 & GG-495 & $1.31 \rightarrow 1.04$  &  1\%  & 12,\,18,\,25  & 1.03 & no \\
T100\,1\,m &       2021/01/03 & 15:40 & 17:47 &   43~ &  120 & 165 & Bessel $R$ & $1.42 \rightarrow 2.82$  &   76\%  & 14,\,24,\,33  & 1.90 & no \\
Cassini\,1.52\,m & 2021/06/29 & 21:15 & 02:21 &  187~ &  60 & 78 & Johnson $V$ & $1.70 \rightarrow 1.03 \rightarrow 1.05$  &  68\%  & 10,\,30,\,55  & 1.83 & no \\
Cassini\,1.52\,m & 2021/06/30 & 22:06 & 02:17 &  181~ &  60 & 78 & Johnson $V$ & $1.35 \rightarrow 1.03 \rightarrow 1.05$  &  58\%  & 10,\,30,\,50  & 1.08 & no \\
\hline
\end{tabular}
}
\end{table*}

Starting from summer 2017, we used an array of five different telescopes (see Table~\ref{tab:obs}) to perform photometric follow-up observations of KELT-16. In particular, we used the Zeiss 1.23\,m telescope at Calar Alto (Spain), the Cassini 1.52\,m telescope at the INAF--OAS Observatory (Italy), the T100 1.0\,m telescope at the TUG Observatory (Turkey), the SPM 84\,cm telescope at the National Astronomical Observatory (Mexico) and the 80\,cm telescope at the OARPAF Observatory (Italy). The observations were carried out through different optical passbands (covering $400-1000$\,nm) to also study the variation of the planetary radius with the wavelength \citep{southworthetal:2012}, see Sect.~\ref{sec:radius_variation}. 

\begin{figure*}
\includegraphics[width=\hsize]{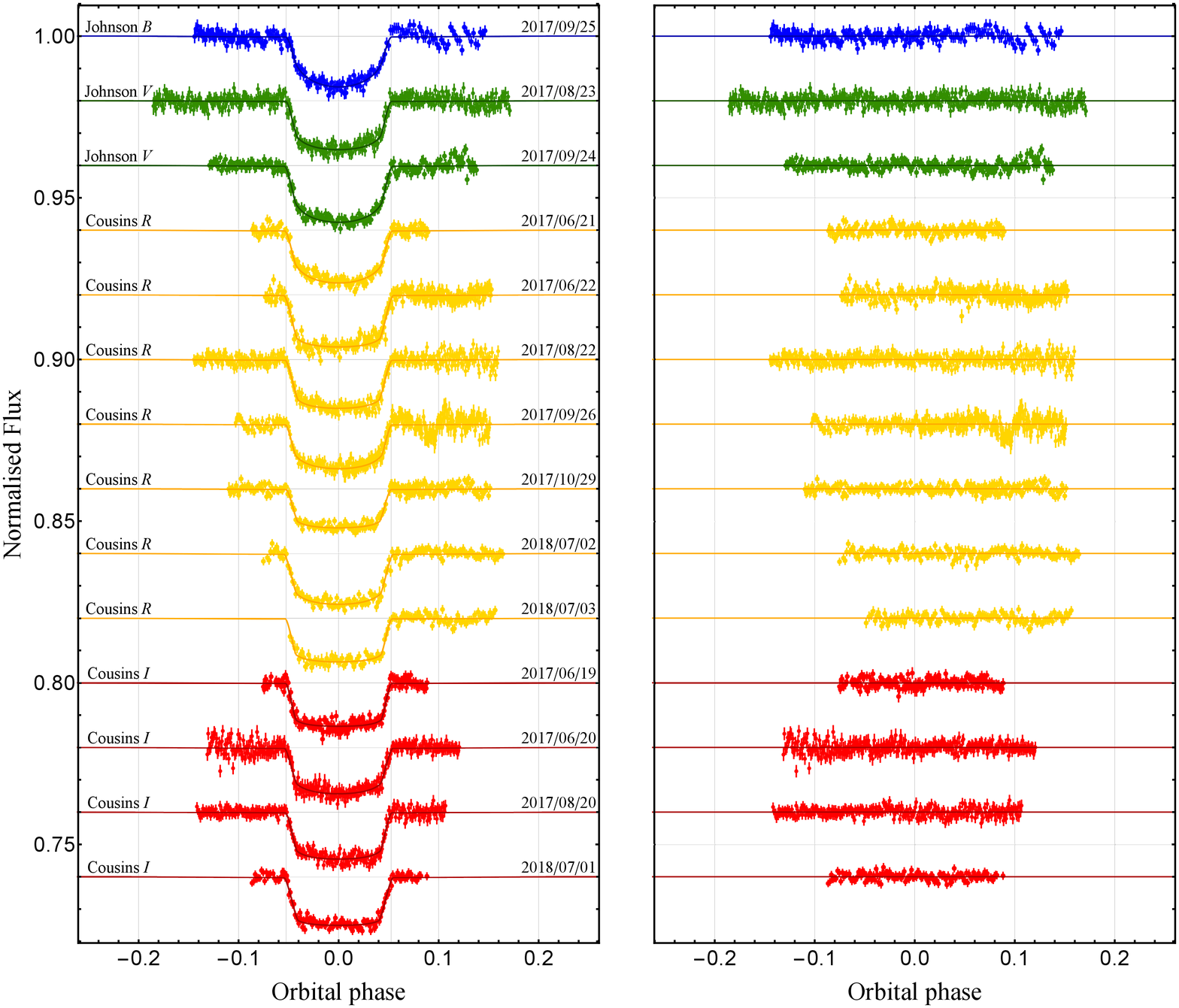}
\caption{{\it Left-hand panel:} light curves of fourteen transits of KELT-16\,b observed with the Zeiss 1.23\,m telescope through four different filters. The labels indicate the observation date and the filter that was used for each observation. They are plotted versus the orbital phase and compared to the best-fitting {\sc jktebop} models. {\it Right-hand panel:} the residuals of each fit.}
\label{fig:CA_lcs}
\end{figure*}

\begin{figure*}
\includegraphics[width=\hsize]{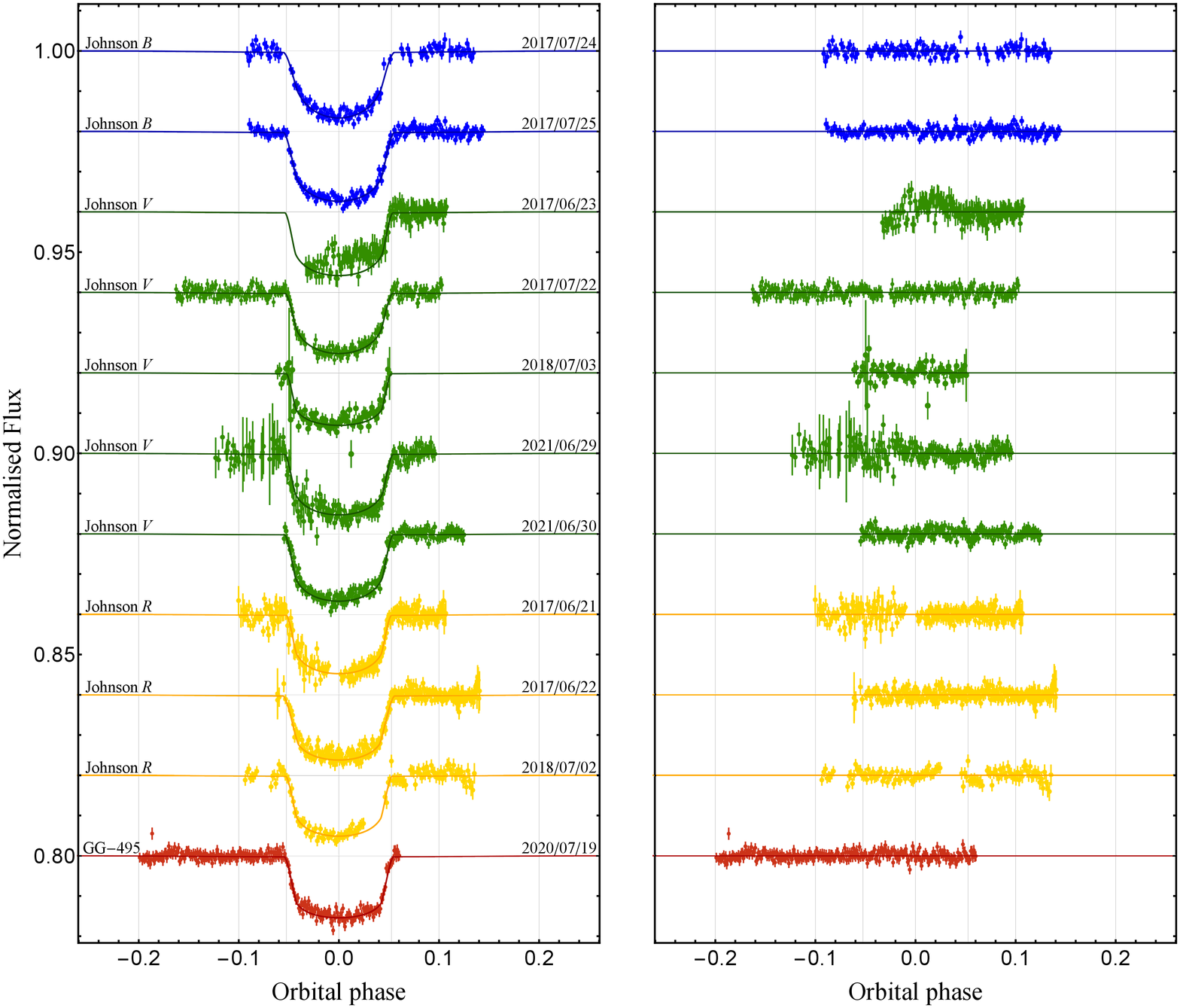}
\caption{{\it Left-hand panel:} light curves of eight transits of KELT-16\,b observed with the Cassini 1.52\,m telescope through four different filters. The labels indicate the observation date and the filter that was used for each observation. They are plotted versus the orbital phase and compared to the best-fitting {\sc jktebop} models. {\it Right-hand panel:} the residuals of each fit.}
\label{fig:Lo_lcs}
\end{figure*}

\begin{figure*}
\includegraphics[width=\hsize]{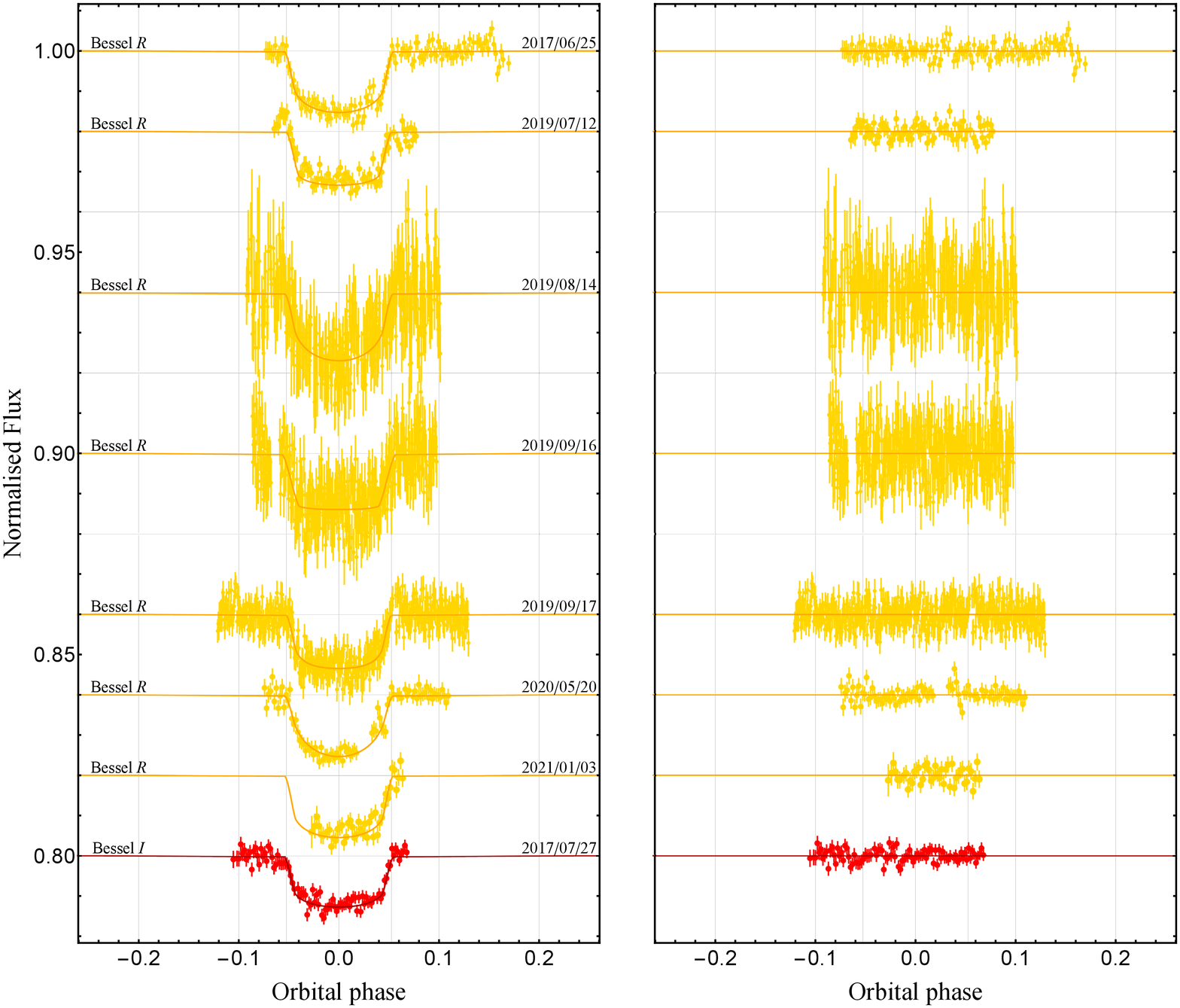}
\caption{{\it Left-hand panel:} light curves of eight transits of KELT-16\,b observed with the T100 1\,m telescope through two different filters. The labels indicate the observation date and the filter that was used for each observation. They are plotted versus the orbital phase and compared to the best-fitting {\sc jktebop} models. {\it Right-hand panel:} the residuals of each fit.}
\label{fig:T100_lcs}
\end{figure*}

\begin{figure}
\includegraphics[width=\hsize]{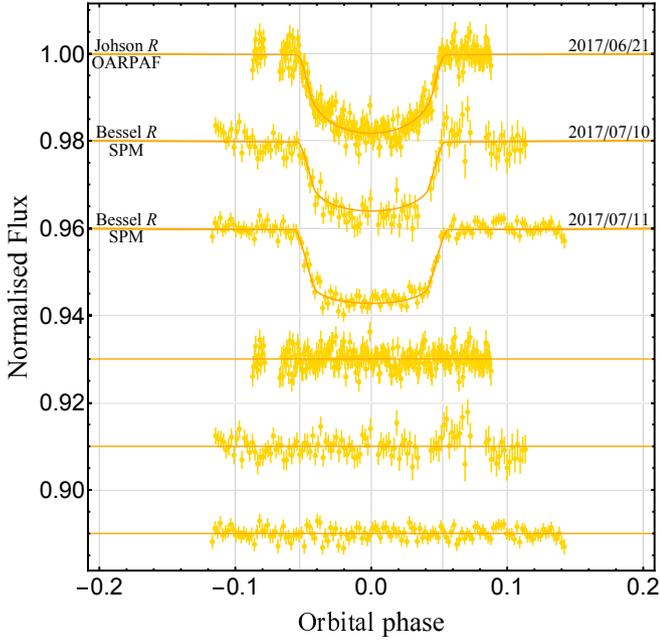}
\caption{Light curves of two transits of KELT-16\,b observed with the SPM 84\,cm telescope and one with the OARPAF 80\,cm telescope. Observations were performed through two different $R$ filters. The labels indicate the observation date and the filter that was used for each observation. They are plotted versus the orbital phase and compared to the best-fitting {\sc jktebop} models. The residuals of the fits are plotted at the base of the figure.}
\label{fig:others_lcs}
\end{figure}

We recorded 30 planetary transits\footnote{The reduced data will be made available at the {\sc cds} catalogue service, {\tt http://cdsweb.u-strasbg.fr/}}, including three transits that were simultaneously observed with two telescopes (Cassini and CA) and one with three telescopes (Cassini, CA and OARPAF), obtaining a total of 36 light curves (Figs.~\ref{fig:CA_lcs}, \ref{fig:Lo_lcs}, \ref{fig:T100_lcs}, \ref{fig:others_lcs}). 

Observations of a planetary-transit event by multiple telescopes from different places are not easy to achieve since one needs the same telescope time at two different observatories and the weather conditions must be favourable at both observation sites. If both conditions are met, the corresponding light curves are useful to better constrain the contact points and, therefore, the mid-transit time, as well as unambiguously understand if any feature present in the light curve is due to a real astrophysical signal  (e.g., starspot) or systematic noise \citep{ciceri:2013,mancini:2013,mancini:2017}. 

The light curves obtained from our simultaneous observations are plotted superimposed in Appendix~\ref{sec:sim_light_curves}. 
\begin{figure*}
\includegraphics[width=\hsize]{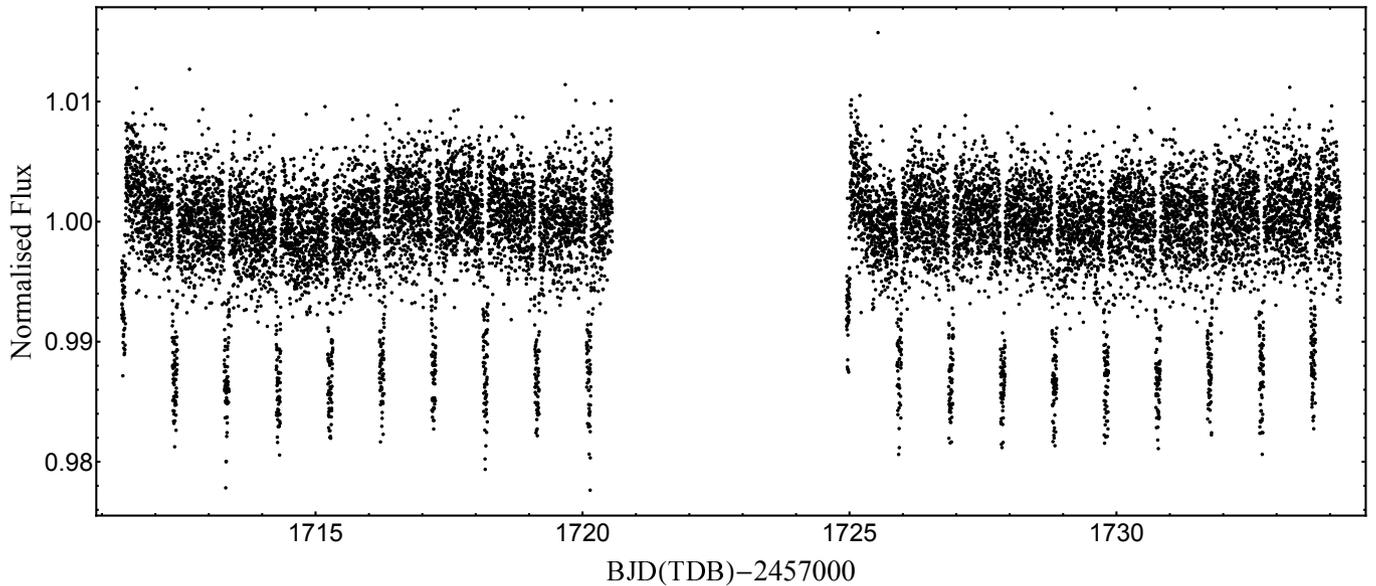}
\caption{The photometric monitoring of KELT-16 by the TESS space telescope.}
\label{fig:Fig_TESS_lc}
\end{figure*}
\begin{figure*}
\includegraphics[width=16cm]{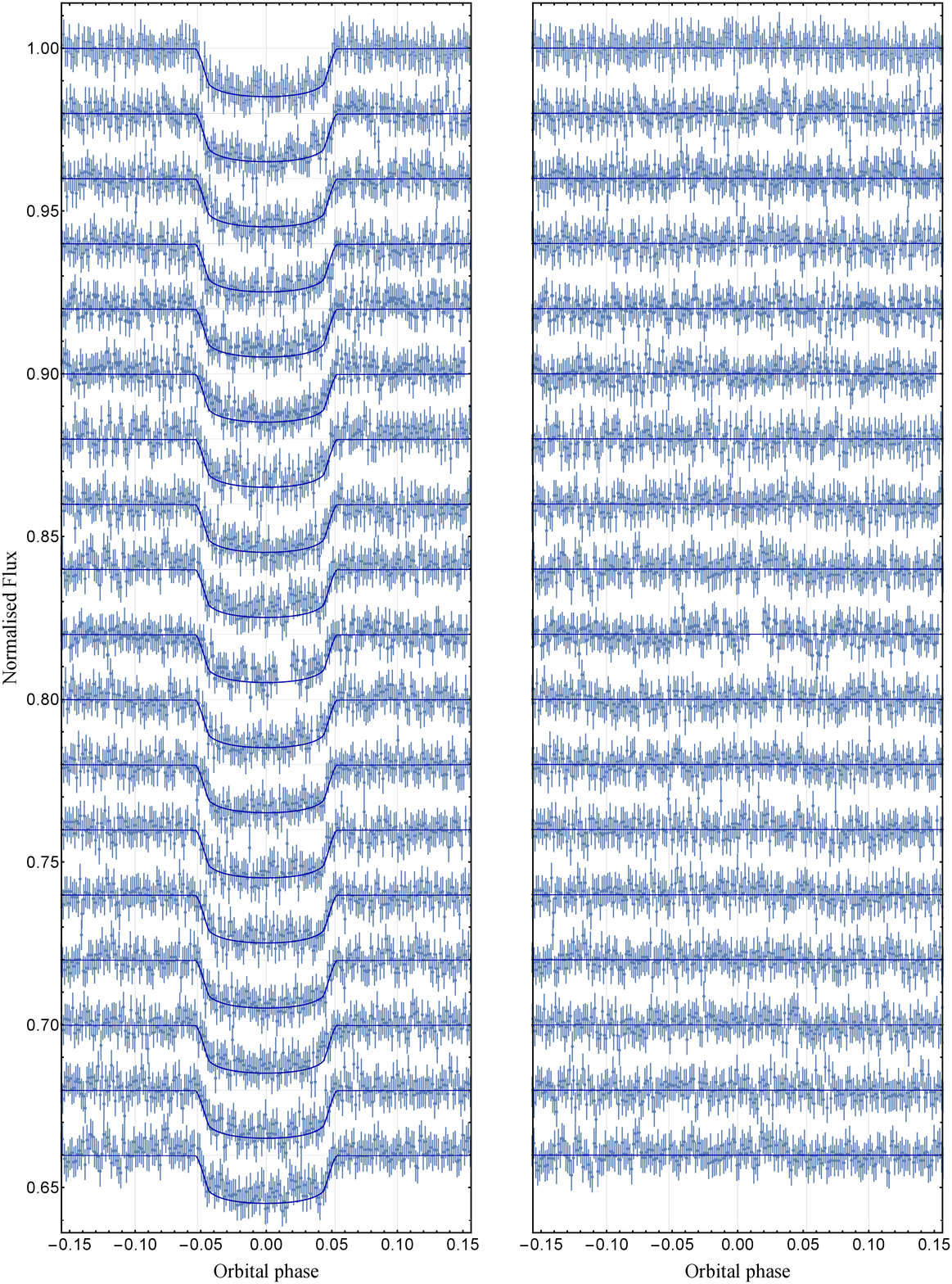}
\caption{{\it Left-hand panel:} TESS light curves of KELT-16\,b transits. They are plotted versus the orbital phase and compared to the best-fitting {\sc jktebop} models. {\it Right-hand panel:} the residuals of each fit.}
\label{fig:Fig_TESS_fit_lc}
\end{figure*}
%

%----------------------------------------
\subsection{CA 1.23\,m telescope}
\label{sec:ca123}
%----------------------------------------
Fourteen complete transits of KELT-16\,b were observed with the Zeiss 1.23\,m telescope at the German-Spanish Astronomical Center at Calar Alto (CA) in Spain. The telescope is equipped with a DLR-MKIII 4k $\times$ 4k camera, which has pixels of size 15\,$\mu$m. Considering that the focal length of the telescope is 9857.1\,mm, the resulting field-of-view (FOV) is 21.5 $\times$ 21.5 arcmin, which guarantees a good number of comparison stars in almost all cases. Observations were all performed using autoguiding and the defocussing method, in order to increase the photometric precision \citep{southworth:2009}. The CCD was windowed to decrease the readout time and increase the cadence of the observations. The transits were observed through standard Johnson $B,\,V$ filters and Cousins $R\,I$ filters. Details of the observations are reported in Table~\ref{tab:obs}.

The data were analysed in the standard way by using the {\sc idl/defot} pipeline \citep{southworth:2009}. In short, we calibrated the raw images of the target with masterbias and masterflat frames, which were obtained by median combining a set of individual bias and sky flat-field images, taken on the same day as each transit observation. We corrected pointing variations by cross-correlating each image against a reference frame, selected considering the airmass. Aperture photometry was performed by placing three apertures on the target and comparison stars. The sizes of the concentric apertures were selected based on the lowest scatter obtained when compared with a fitted model (see Table~\ref{tab:obs}). 
Differential-magnitude light curves were generated for each observing sequence versus an ensemble of comparison stars. A straight line was fitted to the observations outside transit and subtracted to normalize the final light curves to zero differential magnitude. The weights of the comparison stars were simultaneously optimized to minimize the scatter in the data points outside transits. Since the aperture-photometry procedure tends to underestimate the uncertainties of the measurements, they were then scaled so each transit had a reduced $\chi^2$ of $\chi_{\nu}^2=1.0$ versus a best-fitting model calculated with the {\sc jktebop} code (see Sect.~\ref{sec:lc_analysis}). The final light curves are plotted in Fig.\,\ref{fig:CA_lcs}.
Unfortunately, the suboptimal quality of the data is due to the fact that there is only one good comparison star (TYC\,2688-1883-1) in the FOV. This limits the photometric accuracy, even using defocussing, when compared with other transiting systems with similar magnitude, where we have many more comparison stars and we usually obtain lower scatter (<1 mmag) per observation.

%----------------------------------------
\subsection{Cassini 1.52\,m telescope}
\label{sec:cassini}
%----------------------------------------
Eleven transits of KELT-16\,b were observed with the Cassini 1.52\,m telescope from the Astrophysics and Space Science Observatory of Bologna in Loiano (Italy) by using the BFOSC (Bologna Faint Object Spectrograph and Camera). This imager has a back-illuminated CCD with $1300 \times 1340$\,pixels and a pixel size of $20\,\mu$m. The current FOV of this facility is 13 $\times$ 12.6 arcmin. The transits were observed through standard Johnson $B,\,V,\,R$ filters and one with the special uncoated GG-495 glass long-pass filter (transparent at $>500$\,nm), which is used within the EDEN project \citep{gibbs:2020}. Several of these transits were, unfortunately, not well sampled due to bad weather conditions (see Fig.~\ref{fig:Lo_lcs}). Details of the observations are reported in Table~\ref{tab:obs}. The data were reduced and analysed as those from CA (see Sect.~\ref{sec:ca123}).

%----------------------------------------
\subsection{T100 1\,m telescope}
\label{sec:T100}
%----------------------------------------
We observed eight transits of KELT-16\,b with the 1-meter Turkish telescope T100 in T\"UB\.{I}TAK National Observatory of Turkey's (TUG) Bak{\i}rl{\i}tepe Campus, which is located at 2500\,m altitude. The telescope is equipped with a cryo-cooled SI 1100 CCD with $4096\times4096$ pixels, which gives an effective field of view of $20^{\prime}\times20^{\prime}$. We employed a Bessel-$R$ filter in all the observations except that on 2017/07/27, in which a Bessel-$I$ filter was used. We employed the defocusing technique in 5 nights, while we focused the telescope during the nights of August 14, September 16-17 in 2019. Details of the observations are reported in Table~\ref{tab:obs}.
The {\sc AstroImageJ} software package was used \citep{collins:2017} for reducing the CCD images and performing aperture photometry with respect an ensemble of comparison stars. The light curves are plotted in Fig.~\ref{fig:T100_lcs}. 
%----------------------------------------
\subsection{SPM 84\,cm telescope}
\label{sec:SPM}
%----------------------------------------
Two complete transits of KELT-16\,b were observed with the 84\,cm telescope of the Observatorio Astronomico Nacional of San Pedro M\'artir (OAN-SPM), which is located in Northwestern Mexico. This telescope provides a set of instruments that can be mounted according to the observational needs; among these, we have used the Mexman, which is a wide-field imager with a CCD size of $2043\times4612$\,pixels, a resolution of $0.25$\,arcsec\,px$^{-1}$ and a FOV of $8.4\times19.0$\,arcmin. The transits were observed through a Bessel-$R$ filter and the photometric data were reduced and analysed as those from CA (see Fig.~\ref{fig:others_lcs}). Details of the observations are reported in Table~\ref{tab:obs}, while those about this telescope can be found in \citet{ricci:2017}.

%----------------------------------------
\subsection{OARPAF 80\,cm telescope}
\label{sec:OARPAF}
%----------------------------------------
One complete transit of KELT-16\,b was observed with the alt-azimuth OARPAF 80\,cm telescope, located near Mt. Antola in Northern Italy. At the time of the observation, the telescope was equipped with an air-cooled SBIG STL 11000m camera and a standard $UBVRI$ Johnson filter wheel.
The size of this CCD is $2004\times1336$ pixels, with a resolution of $0.29$\,arcsec\,px$^{-1}$ and a FOV of $10\times10$\,arcmin. The transit was observed using the Johnson $R$ filter and the photometric data were reduced and analysed as those from CA (see Fig.~\ref{fig:others_lcs}). Details of the observations are reported in Table~\ref{tab:obs}, while those about this telescope can be found in \citet{ricci:2020}.

%----------------------------------------
\subsection{TESS photomety}
\label{sec:TESS_photomety} %
%----------------------------------------
KELT-16 was observed by TESS \citep{ricker:2015} with the $2$\,min cadence during Sector 15 of its primary mission. Continuous observations of the target star were obtained from 2019-08-15 to 2019-08-25 and from 2019-08-29 to 2019-09-07. Using the \citet{lightkurve:2018} Python package, we downloaded the detrended PDCSAP (Pre-search Data Conditioning Simple Aperture Photometry) light curve, whose known systematics were already removed. The data contains twenty transits of KELT-16\,b, of which eighteen are complete and present out-of-transit data both before the ingress and after the egress, see Fig.~\ref{fig:Fig_TESS_lc}. The suboptimal quality of TESS data is due to the faintness of the KELT-16 star ($V=11.6$\,mag), which is at the limit of the working magnitude of TESS.

%----------------------------------------
\section{Light-curve analysis}
\label{sec:lc_analysis}
%----------------------------------------
%
First of all, we performed a careful analysis of the most suspicious light curves using {\sc prism} and {\sc gemc} codes \citep{tregloan-reed:2013,tregloan-reed:2015}, in order to search for the possible presence of starspots occulted by KELT-16\,b during its transits. It was reported by \citet{tregloan-reed:2019,tregloan-reed:2021} that to fully constrain the physical parameters of a starspot and therefore provide a detection to a high confidence, requires the amplitude of a starspot anomaly to be at least twice the rms scatter of the light curve. Our analysis determined that all the anomaly amplitudes were below this limit and when modelled as a starspot anomaly, gave inconsistent system parameters (i.e. transit depth) compared to the other anomaly free transits, strongly suggesting that the anomalies are non astrophysical systematics.
%This analysis did not reveal any to be present and all the anomaly amplitudes are too low compared to the data %precision. 
The absence of starspots is a conclusion in agreement with a low stellar activity, which is typical for the spectral class of the parent star.

Then, we considered the possible light contribution coming from the companion M3 star (see Sect.~\ref{sec:intro_2}), which is at roughly $0^{\prime \prime}.7$ away from KELT-16, but much fainter and cooler than it. Following \citet{southworth:2010}, we estimated the amount of this ``third light'' by considering the effective temperature of the two stars, the $\Delta K$\,mag measured by \citet{oberst:2017}, the 2MASS $K_{\rm s}$ filter and extrapolated the light ratios to the $B,\,V,\,R,\,I$ bands using synthetic spectra, which were calculated with the ATLAS9 \citep{kurucz:1979,kurucz:1993} and PHOENIX model atmospheres \citep{allard:2001}, and filter-response functions taken from the Isaac Newton Group of Telescopes website\footnote{{\tt http://www.ing.iac.es}.}.
We found the following values for the fraction of light coming from the unresolved star:
\begin{itemize}
\item Johnson $B$: ~~~~~~~~~~~~~~$0.00035 \pm 0.00009$
\item Johnson $V$: ~~~~~~~~~~~~~~$0.00076 \pm 0.00020$
\item Johnson/Cousins $R$:       $0.00147 \pm 0.00037$
\item Johnson/Cousins $I$:\;     $0.00400 \pm 0.00095$
\end{itemize}
Therefore, in the worst case, the contribution of the companion is only $0.4\%$. We made several tests with our $I$-band light curves to see how much effect this is compared to the error bars and found that it does not make a significant difference for the global solution of the system (see Sect.~\ref{sec:physical_properties}).

As a next step, following our consolidated approach (see, for example, \citealt{mancini:2014,southworth:2016} and reference therein), we used the {\sc jktebop} code \citep{southworth:2013} to model the light curves presented in Sect.~\ref{sec:observations}. The orbit of the planet was assumed circular \citep{oberst:2017} and the light curves were modelled using the Levenberg-Marquardt optimization algorithm. In particular, we fitted the sum and ratio of the fractional radii\footnote{The fractional radii are $r_{\star} = R_{\star}/a$ and $r_{\rm p}=R_{\rm p}/a$, where $R_{\star}$ and $R_{\rm p}$ are the true radii of the star and planet, while $a$ is the semi-major axis.} ($r_{\star} + r_{\rm p}$ and $k = r_{\rm p}/r_{\star}$), the orbital period and inclination ($P_{\rm orb}$ and $i$) and the midpoint time of the transit ($T_0$). The limb darkening (LD) of the star was also modelled by applying a quadratic law for describing the LD effect and using the LD coefficients provided by \citet{claret:2004}. We considered the linear coefficient as a free parameter and fixed the non-linear one, which was perturbed during the error-analysis process. Finally, the uncertainties of the fitted parameters were estimated by running a residual-permutation algorithm \citep{southworth:2008}. 

The light curves and their best-fitting models are shown in Figs.~\ref{fig:CA_lcs}, \ref{fig:Lo_lcs}, \ref{fig:T100_lcs}, \ref{fig:others_lcs}. The values of the photometric parameters, which resulted from the fit of each light curve, were combined into weighted means to get the final values. They are reported in Table~\ref{tab:ph_parameters}. 
\begin{table*}
\centering
\caption{Photometric properties of the KELT-16 system derived by fitting the transit light curves with {\sc jktebop}. The values reported in the third column are the weighted means of the results for the individual data sets and are compared with those from other works. }
\label{tab:ph_parameters}
\begin{tabular}{llccc}
\hline
Quantity & Symbol & This work & \citet{oberst:2017} & \citet{maciejewski:2018} \\
\hline
Sum of the fractional radii\dotfill & $r_{\star} + r_{\rm p}$ & $0.33352 \pm 0.00237$ & -- & -- \\ [2pt]
Ratio of the fractional radii\dotfill & $k$ & $0.10814 \pm 0.00087$ & $0.1070_{-0.0012}^{+0.0013}$ & $0.1076 \pm 0.0010$\\ [2pt]
Orbital inclination\dotfill & $i\,(^\circ)$ & $89.72 \pm 0.25$~~ & $84.4_{-2.3}^{+3.0}$ & $84.5_{-1.4}^{+2.0}$ \\ [2pt]
Star's fractional radius\dotfill & $r_{\star}$ & $0.30131 \pm 0.00201$ & $0.310 \pm 0.012$ & $0.3088_{-0.0072}^{+0.0080}$ \\ [2pt]
Planet's fractional radius\dotfill & $r_{\rm p}$ & $0.03264 \pm 0.00032$ & -- & -- \\
\hline
\end{tabular}
\end{table*}

%----------------------------------------
\section{Transit Time Analysis}
\label{sec:transit_time_analysis} %
%----------------------------------------

%----------------------------------------
\subsection{Fitting the timing data}
\label{sec:transit_time_fit} %
%----------------------------------------

\begin{table} 
\centering
\caption{Times of mid-transit for KELT-16\,b and their residuals for a constant period.}
\label{tab:transit_times}
\resizebox{\hsize}{!}{
\begin{tabular}{l r r c} 
\hline  \\[-6pt]%%
~~~Time of minimum      & Cycle~ & ${\rm O}-{\rm C}$~~~ & Reference \\
BJD(TDB)$-2\,400\,000$  & no.~   & (day)~~~             & \\ 
\hline      
$57\,165.85142 \pm 0.00099$ & $ -786.0$ & $-0.00095$ & \citet{oberst:2017}      \\
$57\,166.82179 \pm 0.00086$ & $ -785.0$ & $ 0.00043$ & \citet{oberst:2017}      \\
$57\,168.75660 \pm 0.00185$ & $ -783.0$ & $-0.00275$ & \citet{oberst:2017}      \\
$57\,196.85920 \pm 0.00275$ & $ -754.0$ & $-0.00096$ & \citet{oberst:2017}      \\
$57\,198.79802 \pm 0.00072$ & $ -752.0$ & $-0.00012$ & \citet{oberst:2017}      \\
$57\,228.83690 \pm 0.00100$ & $ -721.0$ & $-0.00004$ & \citet{oberst:2017}      \\
$57\,238.52790 \pm 0.00175$ & $ -711.0$ & $ 0.00103$ & \citet{oberst:2017}      \\
$57\,328.64440 \pm 0.00130$ & $ -618.0$ & $ 0.00114$ & \citet{oberst:2017}      \\
$57\,329.61146 \pm 0.00093$ & $ -617.0$ & $-0.00079$ & \citet{oberst:2017}      \\
$57\,330.58151 \pm 0.00046$ & $ -616.0$ & $ 0.00026$ & \citet{oberst:2017}      \\
$57\,363.52676 \pm 0.00091$ & $ -582.0$ & $-0.00026$ & \citet{oberst:2017}      \\
$57\,714.30206 \pm 0.00071$ & $ -220.0$ & $-0.00054$ & \citet{maciejewski:2018} \\
$57\,914.88456 \pm 0.00051$ & $  -13.0$ & $ 0.00036$ & \citet{patra:2020}       \\
$57\,915.85370 \pm 0.00062$ & $  -12.0$ & $ 0.00050$ & \citet{patra:2020}       \\
$57\,924.57245 \pm 0.00059$ & $   -3.0$ & $-0.00169$ & CA                       \\
$57\,925.54315 \pm 0.00044$ & $   -2.0$ & $ 0.00002$ & CA                       \\
$57\,926.51179 \pm 0.00040$ & $   -1.0$ & $-0.00033$ & CA/Cassini               \\
$57\,927.48150 \pm 0.00026$ & $    0.0$ & $ 0.00039$ & Cassini                  \\
$57\,927.48156 \pm 0.00047$ & $    0.0$ & $ 0.00045$ & \citet{maciejewski:2018} \\
$57\,928.44850 \pm 0.00396$ & $    1.0$ & $-0.00161$ & Cassini                  \\
$57\,930.38847 \pm 0.00061$ & $    3.0$ & $ 0.00038$ & T100                     \\
$57\,945.89131 \pm 0.00184$ & $   19.0$ & $-0.00068$ & SPM                      \\
$57\,946.86111 \pm 0.00110$ & $   20.0$ & $ 0.00013$ & SPM                      \\
$57\,957.51989 \pm 0.00049$ & $   31.0$ & $-0.00001$ & Cassini                  \\
$57\,958.48844 \pm 0.00026$ & $   32.0$ & $-0.00046$ & \citet{maciejewski:2018} \\
$57\,959.45852 \pm 0.00055$ & $   33.0$ & $ 0.00063$ & Cassini                  \\
$57\,960.42762 \pm 0.00050$ & $   34.0$ & $ 0.00073$ & Cassini                  \\
$57\,962.36544 \pm 0.00230$ & $   36.0$ & $ 0.00057$ & T100                     \\
$57\,986.58974 \pm 0.00055$ & $   61.0$ & $ 0.00004$ & CA                       \\
$57\,988.52836 \pm 0.00055$ & $   63.0$ & $ 0.00068$ & CA                       \\
$57\,988.52797 \pm 0.00038$ & $   63.0$ & $ 0.00029$ & \citet{maciejewski:2018} \\
$57\,989.49700 \pm 0.00076$ & $   64.0$ & $ 0.00033$ & CA                       \\
$58\,021.47285 \pm 0.00092$ & $   97.0$ & $-0.00060$ & CA                       \\
$58\,021.47346 \pm 0.00038$ & $   97.0$ & $ 0.00001$ & \citet{maciejewski:2018} \\
$58\,022.44176 \pm 0.00095$ & $   98.0$ & $-0.00068$ & CA                       \\
$58\,022.44219 \pm 0.00046$ & $   98.0$ & $-0.00025$ & \citet{maciejewski:2018} \\
$58\,023.41090 \pm 0.00131$ & $   99.0$ & $-0.00053$ & CA                       \\
$58\,026.31752 \pm 0.00071$ & $  102.0$ & $-0.00090$ & \citet{maciejewski:2018} \\
$58\,056.35704 \pm 0.00064$ & $  133.0$ & $-0.00017$ & CA                       \\
$58\,301.51280 \pm 0.00039$ & $  386.0$ & $ 0.00034$ & CA                       \\
$58\,302.48200 \pm 0.00034$ & $  387.0$ & $ 0.00054$ & CA/Cassini               \\
$58\,303.44940 \pm 0.00065$ & $  388.0$ & $-0.00105$ & CA/Cassini               \\
$58\,334.45858 \pm 0.00061$ & $  420.0$ & $ 0.00035$ & \citet{maciejewski:2018} \\
$58\,365.46578 \pm 0.00063$ & $  452.0$ & $-0.00022$ & \citet{maciejewski:2018} \\
$58\,368.37232 \pm 0.00047$ & $  455.0$ & $-0.00066$ & \citet{maciejewski:2018} \\
$58\,401.31876 \pm 0.00028$ & $  489.0$ & $ 0.00001$ & \citet{maciejewski:2018} \\
$58\,677.48078 \pm 0.00112$ & $  774.0$ & $-0.00097$ & T100                     \\
$58\,710.42937 \pm 0.00167$ & $  808.0$ & $ 0.00186$ & T100                     \\
$58\,711.39561 \pm 0.00071$ & $  809.0$ & $-0.00089$ & TESS                     \\
$58\,713.33461 \pm 0.00049$ & $  811.0$ & $ 0.00013$ & TESS                     \\
$58\,714.30250 \pm 0.00062$ & $  812.0$ & $-0.00098$ & TESS                     \\
$58\,715.27389 \pm 0.00083$ & $  813.0$ & $ 0.00142$ & TESS                     \\
$58\,716.24094 \pm 0.00060$ & $  814.0$ & $-0.00052$ & TESS                     \\
$58\,717.21045 \pm 0.00053$ & $  815.0$ & $-0.00001$ & TESS                     \\
$58\,719.14782 \pm 0.00096$ & $  817.0$ & $-0.00062$ & TESS                     \\
$58\,725.93202 \pm 0.00074$ & $  824.0$ & $ 0.00063$ & TESS                     \\
$58\,726.90076 \pm 0.00076$ & $  825.0$ & $ 0.00038$ & TESS                     \\
$58\,727.86882 \pm 0.00061$ & $  826.0$ & $-0.00056$ & TESS                     \\
$58\,728.83823 \pm 0.00054$ & $  827.0$ & $-0.00014$ & TESS                     \\
$58\,729.80658 \pm 0.00089$ & $  828.0$ & $-0.00079$ & TESS                     \\
$58\,730.77638 \pm 0.00067$ & $  829.0$ & $ 0.00003$ & TESS                     \\
$58\,731.74528 \pm 0.00090$ & $  830.0$ & $-0.00007$ & TESS                     \\
$58\,732.71525 \pm 0.00061$ & $  831.0$ & $ 0.00091$ & TESS                     \\
$58\,733.68336 \pm 0.00077$ & $  832.0$ & $ 0.00003$ & TESS                     \\
$58\,743.37244 \pm 0.00092$ & $  842.0$ & $-0.00082$ & T100                     \\
$58\,744.34073 \pm 0.00062$ & $  843.0$ & $-0.00153$ & T100                     \\
$58\,990.46559 \pm 0.00139$ & $ 1097.0$ & $-0.00085$ & T100                     \\
$59\,050.54472 \pm 0.00035$ & $ 1159.0$ & $ 0.00072$ & Cassini                  \\                       
$59\,218.17959 \pm 0.00080$ & $ 1332.0$ & $-0.00016$ & T100                     \\ 
$59\,395.50478 \pm 0.00062$ & $ 1515.0$ & $-0.00064$ & Cassini                  \\
$59\,396.47467 \pm 0.00046$ & $ 1516.0$ & $ 0.00026$ & Cassini                  \\
\hline
\end{tabular}
}
\end{table}
\begin{figure*}
\includegraphics[width=16cm]{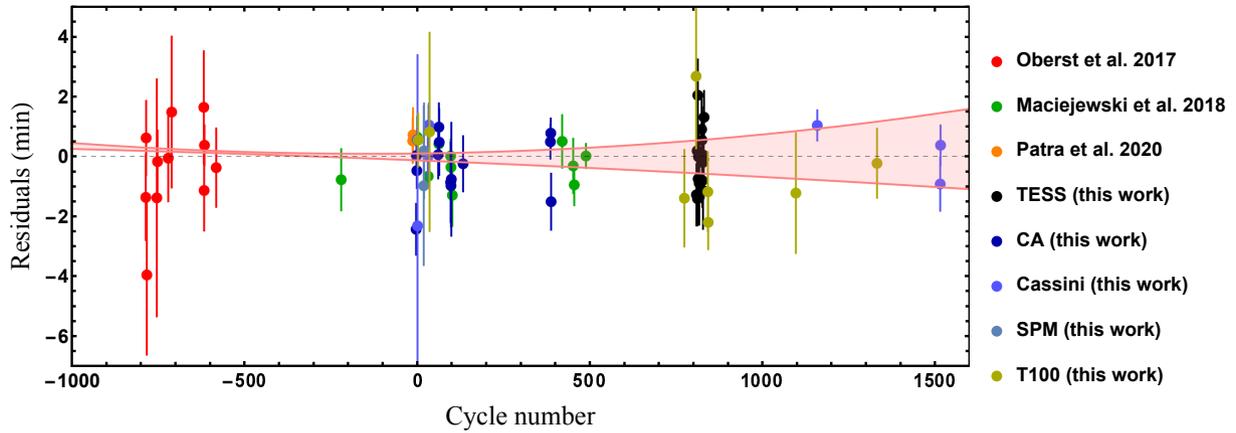}
\caption{Plot of the residuals of the timings of mid-transit of KELT-16\,b versus a linear ephemeris. The points are coloured based on their source (see the legend). CA points (dark blue) include the times averaged on simultaneous observations (see Table~\ref{tab:obs} and Appendix~\ref{sec:sim_light_curves}). The red band represents $1\,\sigma$ uncertainty on the orbital decay model. Zooms in to the two best-sampled regions are shown in Fig.~\ref{fig:zoom}.}
\label{fig:Fig_O-C}
\end{figure*}
\begin{table*} 
\centering
\caption{Summary of the best-fitting values of the parameters obtained by using a linear-, a quadratic and a cubic-ephemeris model for the KELT-16\,b time residuals. The corresponding reduced chi-square, AIC, BIC and root-mean-square-deviation (rmsd) values are also reported.}
\label{tab:transit_times-best-fits}
\resizebox{\hsize}{!}{
\begin{tabular}{l c c c c c c c c} 
\hline  \\[-6pt]%
Poly order & $\chi_{\nu}^2$  & $T_0$\,(BJD$_{\rm TDB}-2400000$) & Orbital Period (day) & Quadratic term & Cubic term & AIC & BIC & rmsd\,(s) \\
\hline      
Linear    & 0.98 & $57927.481100 \pm 0.000080$ & $0.968993061 \pm 0.000000139$ & -- & -- & 73.29 & 77.82 & 67.68 \\
Quadratic & 0.97 & $57927.481114 \pm 0.000082$ & $0.968993190 \pm 0.000000211$ & $-(1.62 \pm 2.01)\times 10^{-10}$ & -- & 74.64 & 81.43 & 66.49 \\
Cubic     & 0.96 & $57927.481145 \pm 0.000095$ & $0.968993123 \pm 0.000000236$ & $-(3.49 \pm 3.54)\times 10^{-10}$ & $(1.81 \pm 2.82)\times 10^{-13}$ & 76.23 & 85.28 & 65.97 \\
\hline
\end{tabular}
}
\end{table*}
From the study of our new light curves (see Sect.~\ref{sec:lc_analysis}), we can estimate the mid-transit time for each of the transits that we recorded. However, we have to consider that time-correlated errors (i.e., red noise) can significantly affect ground-based data and, therefore, the determination of transit times. In order to derive more realistic uncertainties for each point in our light curves, we assessed the red noise by using the $\beta$ approach (e.g., \citealt{gillon:2006,winn:2008,gibson:2008,nikolov:2012,southworthetal:2012a,mancini:2013b}. Practically, we inflated the error bars further by multiplying the data weights by a factor $\beta \ge 1$, which is a measurement of how close the data noise is to the Poisson approximation. The $\beta$ factor is found by binning the data and determining the ratio between the size of the residuals versus what would be expected if the data followed the Poisson distribution. We evaluated the values of the $\beta$ factor for between two and ten data points for each light curve, and adopted the largest $\beta$ value.

We then estimated the mid-transit time for each of the transits that we recorded by running the {\sc jktebop} code another time. In particular, concerning the four transits that we simultaneously observed with two telescopes (see Appendix~\ref{sec:sim_light_curves}), we obtained the following values for the mid-transit times:
\begin{center}
\begin{tabular}{lc}
 \hline
 Telescope & $T_{\rm mid}$ (BJD$_{\rm TDB}$)\\ %
 \hline
{\bf 2017.06.21} \\
CA\,1.23\,m & $t_1=2\,457\,926.51170 \pm 0.00048$ \\ 
Cassini\,1.52\,m & $t_2=2\,457\,926.51199 \pm 0.00071$ \\
OARPAF\,80\,cm & $t_3=2\,457\,926.51419 \pm 0.00052$ \\ [2pt]
& $t_2-t_1=(25.1 \pm 102.8)$\, sec  ~~~~  0.3-$\sigma$ \\
& $t_3-t_1=(215.1 \pm 106.3)$\, sec ~~  3.5-$\sigma$ \\
& $t_3-t_2=(190.1 \pm 86.4)$\, sec ~~~~ 1.5-$\sigma$ \\[4pt] 
{\bf 2017.06.22} \\
CA\,1.23\,m & $t_1=2\,457\,927.47825 \pm 0.00047$ \\
Cassini\,1.52\,m & $t_2=2\,457\,927.48150 \pm 0.00026$ \\ [2pt]
& $t_2-t_1=(280.8 \pm 63.1)$\, sec ~~~~ 6.1-$\sigma$ \\ [4pt]
{\bf 2018.07.02} \\ 
CA\,1.23\,m & $t_1=2\,458\,302.48199 \pm 0.00037$ \\
Cassini\,1.52\,m & $t_2=2\,458\,302.48202 \pm 0.00087$ \\ [2pt]
& $t_2-t_1=(2.6 \pm 107.1)$\, sec ~~~~ 0.03-$\sigma$ \\ [4pt]
{\bf 2018.07.03} \\
CA\,1.23\,m & $t_1=2\,458\,303.44916 \pm 0.00093$ \\
Cassini\,1.52\,m & $t_2=2\,458\,303.44984 \pm 0.00091$ \\ [2pt]
& $t_2-t_1=(58.8 \pm 159.0)$\, sec ~~~~ 0.5-$\sigma$ \\ 
 \hline
\end{tabular}
\end{center}
We noticed that when the transit is simultaneously and completely monitored by two telescopes, including the out-of-transit phases, the difference between the mid-transit times is minimal ($\sim 2.6$\,sec, with 0.03-$\sigma$). Instead, when the transit is not completely monitored by the two telescopes, then the difference is greater ($\sim 215$\,sec and $\sim 280$\,sec in our worst cases, i.e. a difference of 3.5-$\sigma$ and 6.1-$\sigma$, respectively). 
This happening was already noted by \citet{barros:2013}. This stresses how important it is to get complete light curves (including the out-of-transit points) of transit events in order to achieve reliable estimates of the mid-transit times. Considering the differences in the above table, we decided to reject the timing from OARPAF\,80\,cm taken on 2017.06.21 and that from CA taken on 2017.06.22. For the remaining timings of the table, as in previous works of our series, we decide to take the weighted mean of the times for each of these four transits. 

Finally, we assembled a final list of 69 transit timings by joining our measurements with those from \citet{oberst:2017}, \citet{maciejewski:2018} and \citet{patra:2020}. They are reported in Table~\ref{tab:transit_times}. 

We selected the reference epoch as that corresponding to the first observation of our campaign (see Fig.~\ref{fig:Fig_O-C}) and we made several attempts to fit the above-mentioned list of timing data. In particular, we tried both a linear and a quadratic ephemeris in the forms
\begin{equation}
\label{eq:fit_linear_1}
T_{\rm mid}=T_0+P_{\rm orb}E\,,
\end{equation}
\begin{equation}
\label{eq:fit_quadratic_1}
T_{\rm mid}=T_0+P_{\rm orb}E+\frac{1}{2}\frac{dP_{\rm orb}}{dE}E^2\,. 
\end{equation}
As usual, $E$ represents the number of orbital cycles after the reference epoch $T_{0}$, while $\frac{dP_{\rm orb}}{dE}$ is the change in the orbital period between succeeding transits. The fit of the mid-transit times with a straight line gave the following refinement of the linear transit ephemerides
\begin{equation}
\label{eq:fit_linear_2}
T_{\rm mid}={\rm BJD_{TDB}}\,2\,457\,927.481100\,(80)+ 0.968993061\,(139)\,E\,,
\end{equation}
with a reduced chi-square of $\chi_{\nu}^2=0.98$ and a root-mean-square deviation (rmsd) scatter of 67.68\,s (the quantities in brackets represent the uncertainties in the preceding digits). Instead, the best-fitting quadratic ephemeris resulted to be
\begin{equation}
\label{eq:fit_quadratic_2}
\begin{split}
T_{\rm mid}={\rm BJD_{TDB}}\,2\,457\,927.481114\,(82) + 0.968993190\,(211)\,E + \\ 
-(1.62 \pm 2.01)\times 10^{-10} E^2 \,, ~~~~~~~~~~~~~~~ \\
\end{split}
\end{equation}
with a slightly lower $\chi^2_{\nu}$ and rmsd scatter, i.e. 0.97 and 66.24\,s, respectively.
We also tried a fit with a cubic ephemeris and, also in this case, we found a negligible improvement compared to the previous cases; see Table~\ref{tab:transit_times-best-fits}, where we summarise the results of our analysis.

Since the $\chi^2_{\nu}$ of the three models was very similar to each other, we estimate the Akaike Information Criterion (AIC) and the Bayesian Information Criterion (BIC). Both these criteria slightly prefer the linear model over the quadratic and cubic models (see Table~\ref{tab:transit_times-best-fits}). The timing residuals from the linear ephemeris are plotted in Fig.~\ref{fig:Fig_O-C} together with the quadratic model.

Following a suggestion of the referee, we excluded the light curves that have missing ingress/egress or strong systematics and repeated the analysis of the transit times without these values to check if the results are consistent. In particular, we excluded the light curve recorded by the Cassini 1.52\,m telescope on 2017/06/23 because the ingress is missing; we excluded the two T100 light curves (recorded on 2019/08/14 and 2019/09/16) because they were observed without using defocussing; we excluded the T100 light curve recorded on 2021/01/03 because the ingress is missing. Having excluded these four timings, we remade the TTV analysis and did not find significant differences compared to the case with the entire dataset.

%----------------------------------------
\subsection{Orbital-decay analysis}
\label{sec:orbital_decay} %
%----------------------------------------
The analysis presented in the Sect.~\ref{sec:transit_time_fit} shows that both linear and quadratic models fit the mid-transit times of KELT-16\,b equally well. The impasse can only be overcome by acquiring new planetary-transit measurements. The quadratic terms can be progressively constrained to be smaller and smaller as more data are added. Based on the current data and following the approach of previous studies (e.g., \citealt{maciejewski:2016,patra:2017,southworth:2019}), we found that the change in the orbital period is 
$\frac{dP_{\rm orb}}{dE}=-(3.2 \pm 4.0)\times 10^{-10}$\,days per orbital cycle and, therefore, the period derivative is
$\dot{P}_{\rm orb}=\frac{1}{P_{\rm orb}}\frac{dP_{\rm orb}}{dE}=-10.6 \pm 13.1$\;ms\,yr$^{-1}$, 
consistent with a constant orbital period or with an orbital period that shrink to zero in a time larger than $8$\,Myr.

The rate of the orbital decay, which we have deduced, may be used to limit the modified tidal dissipation quality factor, via Eq.~(\ref{eq:quality-factor}), to $Q^{\prime}_{\star}>(1.9 \pm 0.8) \times 10^5$; this result is based on the $95\%$ confidence lower limits on $\dot{P}_{\rm orb}$, while the uncertainties come from propagating the errors in $M_{\rm p}/M_{\star}$ and $R_{\star}/a$ (see Table~\ref{tab:finalparameters}).

%----------------------------------------
\section{Physical properties}
\label{sec:physical_properties}
%----------------------------------------
%----------------------------------------
\subsection{Analysis of stellar parameters}
\label{sec:barbato}
%----------------------------------------
We reviewed the stellar parameters of the KELT-16 star by fitting its Spectral Energy Distribution (SED) using the MESA Isochrones and Stellar Tracks (MIST; \citealt{dotter:2016,choi:2016}) through the \texttt{EXOFASTv2} suite \citep{eastman:2019}. To be precise, following \citet{barbato:2020}, we fitted the stellar magnitudes \citep{oberst:2017}, imposing Gaussian priors on the effective temperature and the metallicity. These priors were based on the spectroscopic measurements and the parallax obtained from Gaia DR3. From this analysis, we estimated the stellar atmospheric parameters, which resulted to be in good agreement with those of the discovery paper, see Table~\ref{tab:finalparameters}.

%----------------------------------------
\subsection{Physical parameters of the planetary system}
\label{sec:Physical_parameters}
%----------------------------------------
\begin{table*} 
\centering
\caption{Physical parameters of the planetary system KELT-16 derived in this work (Sect.~\ref{sec:Physical_parameters}), compared with those from other works. Where two error bars are given, the first refers to the statistical uncertainties, while the second to the systematic errors. {\bf Notes}. $^{a}$ This value was obtained from the SED fitting procedure (Sect.~\ref{sec:barbato}). $^{b}$ This value was obtained from the transit-time analysis (Sect.~\ref{sec:transit_time_analysis}).
}
\label{tab:finalparameters}
\resizebox{\hsize}{!}{
\begin{tabular}{l c c c c c c} \hline
Quantity & Symbol & Unit & This work & \citet{oberst:2017} & \citet{maciejewski:2018} & \citet{patra:2020}\\
\hline  \\[-6pt]%%
\multicolumn{1}{l}{\textbf{Stellar parameters}} \\
Stellar effective temperature$^{a}$ \dotfill & $T_{\rm eff}$ & K & $6237^{+55}_{-53}$ & $6236 \pm 54$ & -- & --             \\ [2pt]
Stellar metallicity$^{a}$ \dotfill & [Fe/H] & dex & $-0.006^{+0.082}_{-0.082}$ & $-0.002^{+0.086}_{-0.085}$ & -- & --       \\ [2pt]
Stellar mass        \dotfill & $M_{\star}$ & $M_{\sun}$ & $1.195 \pm 0.037 \pm 0.024$ & $1.211_{-0.046}^{+0.043}$ & -- & -- \\ [2pt]
Stellar radius      \dotfill & $R_{\star}$ & $R_{\sun}$ & $1.315 \pm 0.016 \pm 0.009$ & $1.360_{-0.053}^{+0.064}$ & -- & -- \\ [2pt]
Stellar surface gravity \dotfill & $\log g_{\star}$ & cgs & $4.278 \pm 0.007 \pm 0.003$ & $4.253_{-0.036}^{+0.031}$ & -- & -- \\ [2pt]
Stellar density     \dotfill & $\rho_{\star}$ & $\rho_{\sun}$ & $0.5256 \pm 0.0100$ & $0.481_{-0.057}^{+0.056}$ & -- & --   \\ [2pt]
Age           \dotfill & $\tau$          & Gyr   & $3.0_{-0.5\,-0.4}^{+0.7\,+0.4}$   & $3.1 \pm 0.3$ & -- & --              \\ [2pt]
$V$-band extinction$^{a}$ \dotfill & $A_{V}$ & mag & $0.179^{+0.055}_{-0.054}$ & $0.04 \pm 0.04$ & -- & --                   \\ [2pt]
Parallax$^{a}$ & $\varpi$ & mas & $2.247^{+0.013}_{-0.013}$ & -- & -- & --                                                  \\ [2pt]
Distance$^{a}$ & $d$ & pc & $445.0 \pm 2.6$ & $399 \pm 19$ & -- & --                                                         \\ [2pt]
\hline \\[-6pt]%
\multicolumn{1}{l}{\textbf{Planetary parameters}} \\
Planetary mass   \dotfill & $M_{\rm p}$ & $M_{\rm Jup}$ & $2.71  \pm 0.15  \pm 0.04$  & $2.75_{-0.15}^{+0.16}$ & -- & --    \\ [2pt]
Planetary radius \dotfill & $R_{\rm p}$ & $R_{\rm Jup}$ & $1.383 \pm 0.023 \pm 0.009$ & $1.415_{-0.067}^{+0.084}$ & -- & -- \\ [2pt]
Planetary surface gravity \dotfill & $g_{\rm p}$ & m\,s$^{-2}$  & $35.1 \pm 2.0$ & $33.9_{-3.6}^{+3.4}$ & -- & --           \\ [2pt]
Planetary density \dotfill & $\rho_{\rm p}$ & $\rho_{\rm Jup}$ & $0.958 \pm 0.062 \pm 0.006$ & $0.90 \pm 0.14$ & -- & --    \\ [2pt]
Equilibrium temperature  \dotfill & $T_{\rm eq}$ & K & $2417 \pm 22$ & $2453_{-47}^{+77}$ & -- & --                         \\ [2pt]
Safronov number  \dotfill & $\Theta$ & & $0.0667 \pm 0.0036 \pm 0.0004$ & $0.0654 \pm 0.0045$ & -- & --                     \\
\hline \\[-6pt]%
\multicolumn{1}{l}{\textbf{Orbital parameters}} \\
Semi-major axis \dotfill & $a$ & au & $0.02035 \pm 0.00021 \pm 0.00014$ & $0.02044_{-0.00026}^{+0.00024}$ & -- & -- \\ [2pt]
Inclination     \dotfill & $i$ & degree  & $89.72   \pm 0.14   $ & $84.4_{-2.3}^{+3.0}$ & $84.5^{+2.0}_{-1.4}$ & -- \\ [2pt] %
\multicolumn{1}{l}{\emph{~~~~~Constant period}} \\
Time of mid-transit$^{b}$ \dotfill & $T_{0}$  & BJD$_{\rm TDB}$ & $2\,457\,927.481100\,(80)$ & $2\,457\,165.85179\,(49)$ & $2\,457\,247.24774\,(24)$ &  $2\,457\,910.03913\,(11)$ \\
Period$^{b}$ \dotfill & $P_{\rm orb}$ & days & $0.96899306\,(14)$ & $0.9689951\,(24)$ & $0.96899320\,(29)$ & $0.96899319\,(30)$ \\
\multicolumn{1}{l}{\emph{~~~~~Orbital decay}} \\
Time of mid-transit$^{b}$ \dotfill & $T_{0}$  & BJD$_{\rm TDB}$ & $2\,457\,927.481114\,(82)$ & -- & -- & $2\,457\,910.03918\,(15)$ \\
Period$^{b}$ \dotfill & $P_{\rm orb}$ & days & $0.96899319\,(21)$ & -- & -- & $0.96899314\,(33)$ \\
\hline
\end{tabular}
}
\end{table*}

The main physical properties of the KELT-16 system were determined using our robust ``Homogeneous Studies'' approach (see for example \citealt{southworth:2012,mancini:2013} and references therein), which makes use of the photometric parameters reported in Table~\ref{tab:ph_parameters}, the stellar radial-velocity amplitude $K_{\star}= 494 \pm 25$m\,s$^{-1}$ \citep{oberst:2017}, the spectroscopic parameters (Sect.~\ref{sec:barbato}) and a set of theoretical stellar models. We calculated the properties of the system by using standard formulae, including the velocity amplitude of the planet, $K_{\rm p}$. Then, we iteratively adjusted the value of $K_{\rm p}$ to find the best agreement between the values of $r_{\star}$ and $T_{\rm eff}$ that we measured from the observations and those predicted by a single theoretical model of the calculated mass. This procedure was repeated for a wide grid of ages and five different theoretical models.

Finally, we took the unweighted mean of the five sets of values as the final set of physical properties of the system. They are reported in Table~\ref{tab:finalparameters}. The systematic uncertainties of these values were calculated as the standard deviation of the results from the five models for each output parameter. Table~\ref{tab:finalparameters} also shows the physical properties found by other authors. Our results are in good agreement with them but more precise because of the much more extensive photometry of the system presented in the current work.

%----------------------------------------
\subsection{Variation of the planetary radius with the wavelength}
\label{sec:radius_variation} %
%----------------------------------------
It has been well ascertained that the transmission spectra of hot Jupiters show characteristic absorption features at particular wavelengths (see, e.g., \citealt{sing:2018}). In particular, those for which the incident stellar flux is $>10^9$\,erg\,s$^{-1}$\,cm$^{-2}$ are expected to host a large amount of absorbing substances in their atmospheres, such as gaseous titanium oxide (TiO) and vanadium oxide (VO), acting in a window of the optical region between 450 and 700\,nm \citep{fortney:2008}. Using the multi-colour light curves that we took through different passbands (i.e. $B,\,V,\,R$\footnote{Even though different $R$ filters were used, they are very similar considering their bandwidth and effective wavelength.}, $I$), we attempted to reconstruct a low-resolution transmission spectrum of KELT-16\,b. We excluded from this analysis the low-quality and the incomplete light curves that we presented in Sect.~\ref{sec:observations} (also see Table~\ref{tab:obs} and Figs~\ref{fig:CA_lcs}, \ref{fig:Lo_lcs} and \ref{fig:others_lcs}), because from their fits we can get inaccurate value of the transit depth. 
We also excluded the TESS light curves because the long-pass filter of TESS is just too wide ($>500$\,nm) for our purposes. 

\begin{figure}
\includegraphics[width=\hsize]{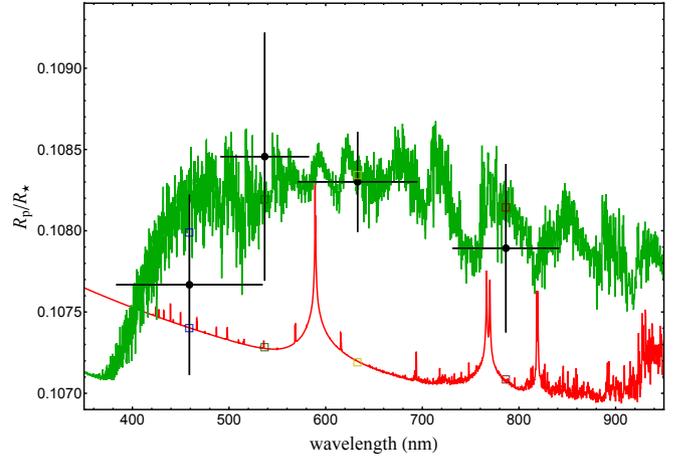}
\caption{Variation of the planetary radius, in terms of the planet/star radius ratio, with wavelength. The points are from the ground-based transit observations presented in this work. The vertical bars represent the errors in the measurements and the horizontal bars show the FWHM transmission of the passbands used. The observational points are compared with two synthetic spectra from \citet{fortney:2010} (see text for details). Coloured squares represent band-averaged model radii over the bandpasses of the observations.}
\label{fig:Fig_radius_variation}
\end{figure}

Following a general approach, we run again {\sc jktebop} for each of our light curves to calculate the ratio of the radii in each passband, fixing this time the other photometric parameters to their best-fitting values that we previously estimated (see Table~\ref{tab:finalparameters}). This returned a set of $k=R_{\rm p}/R_{\star}$ values which are directly comparable and whose error bars exclude common sources of uncertainty. Then, we made a weighted mean of these values for each of the four passbands, obtaining the following: $k_{B}=0.10767 \pm 0.00055$, $k_{V}=0.10846 \pm 0.00076$, $k_{R}=0.10825 \pm 0.00112$ and $k_{I}=0.10833 \pm 0.00051$. These values are shown in Fig~\ref{fig:Fig_radius_variation} and compared with two one-dimensional
model atmospheres, which were obtained by \citet{fortney:2010}. In particular, the red line has been calculated for Jupiter's gravity (25\,m\,s$^{-2}$) with a base radius of $1.25\,R_{\rm Jup}$ at the 10\,bar level and at 2500\,K. The opacity of TiO and VO molecules is excluded from the model and the optical transmission spectrum is dominated by H$_2$/He Rayleigh scattering in the blue, and pressure-broadened neutral atomic lines of sodium at 589\,nm and potassium at 770\,nm. The green line represents a model similar to the previous one, but the opacity of TiO and VO molecules was included.
This model shows significant optical absorption that broadly peaks around 700 nm, with a sharp fall-off in the blue and a shallower fall-off in the red. Unfortunately, the accuracy of our data is not at the same level of accuracy as the atmosphere models; even though they indicate a possible radius variation between wavelength ranges $350-420$\,nm and $500-800$\,nm, practically, they show a flat transmission spectrum to within the experimental uncertainties. 

%----------------------------------------
\subsection{Flux ratio of KELT-16\,b from TESS phase curve}
\label{sec:phase_curve} %
%----------------------------------------
Due to its short period, KELT-16\,b has been selected as a back-up target for the JWST Early Release Science program \citep{bean:2018}. This because the time required to observe an its full orbit phase curve is relatively small when compared with a more typical hot Jupiter ($P_{\rm orb} \sim 3$\,days). Moreover, KELT-16\,b is highly irradiated and, therefore, will give a large thermal emission signal.

\begin{figure}
\includegraphics[width=\hsize]{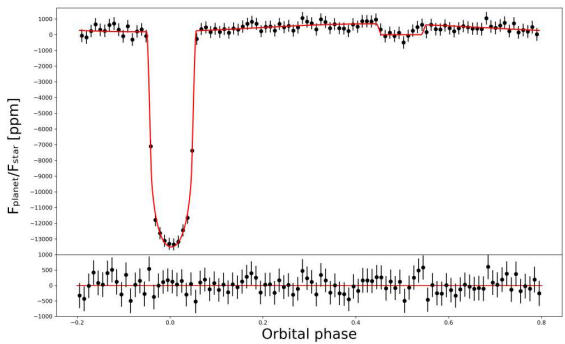}
\caption{{\it Top panel}: phase curve of KELT-16 from TESS data, fitted with the Starry package. {\it Bottom panel}: residual of the fit.} 
\label{fig:Fig_phasecurve}
\end{figure}

By using the SPOC pipeline, we extracted the data of KELT-16 from the TESS database and generated a phase curve with the periodicity and time of transit shown in Table~\ref{tab:finalparameters}. We exploited the Starry python package \citep{luger:2019} to fit both the transit and the secondary eclipse to measure the flux ratio of the planet over its star, see Fig~\ref{fig:Fig_phasecurve}. 
We approximate KELT-16\,b to be tidally-locked and model it with a simple dipole brightness map, where the bright side is facing the star, but we also take into account a possible offset that captures the eventual misalignment between the hot spot of the planet and the sub-stellar point, which causes the maximum of the planet flux to happen outside of the secondary eclipse. The mass and radius of both the star and the planet are set to be normally distributed around the value and standard deviation from Table~\ref{tab:finalparameters}, while the offset and the amplitude of the planet flux are free to vary within reasonable limits. A brightness temperature for both the day side and the night side of the planet is then computed with the Planck's law. 
The results are shown in Table~\ref{tab:posterior_values} and the distributions are plotted in Fig.~\ref{fig:Fig_distribution}, from where result an offset of $25 \pm 14$\,$^{\circ}$ and a $F_{\rm p}/F_{\star}$ of $434 \pm 42$\,ppm. Finally, the brightness temperature of the day- and night-side simple model are $T_{\rm day}=3190 \pm 61$\,K and $T_{\rm night}=2668 \pm 56$\,K, respectively. These values are compatible within $1\,\sigma$ and $2\,\sigma$, respectively, of those found by \citet{bell:2021} with Spitzer measurements.
\begin{table}
\centering
\caption{Posterior values and their standard deviations over the model parameters from a Starry computation with 8000 draw iterations. The day and the night flux are defined as the fluxes ratio that the planet emits over the star's respectively at phase 0.5 and 0.0 in ppm (also the average planet amplitude is in ppm). hdi\_$3\%$ and hdi\_$97\%$ represent the lower and upper bounds of a $\sim95\%$ credible interval, which contains the true parameter value with $\sim95\%$ probability.}
\label{tab:posterior_values}
\resizebox{\hsize}{!}{
\begin{tabular}{l c c c c | c}
\hline
Quantity & Unit & This work & hdi\_$3\%$ & hdi\_$97\%$ & \citet{bell:2021} \\
\hline  \\[-6pt]%
$F_{\rm p}/F_{\star}$\dotfill & (ppm)        & $434.19 \pm 41.61$ & 361.08 & 519.88 & $4810_{-310}^{+330}$ \\ [2pt]
Max flux offset\dotfill       &  $^{\circ}$E & $ 25.25 \pm 14.03$ &  $-0.56$ &  52.50 & $ -38_{-15}^{+16}$   \\ [2pt]
Day flux\dotfill              & (ppm)        & $654.61 \pm 71.18$ & 521.35 & 786.51 & -- \\[2pt]
Night flux\dotfill            & (ppm)        & $213.78 \pm 31.94$ & 158.69 & 276.11 & -- \\[2pt]
$T_{\rm day}$\dotfill         & K            & $3190 \pm 61$ & $3077$ & $3301$ & $3070_{-150}^{+160}$ \\ [2pt]
$T_{\rm night}$\dotfill       & K            & $2668 \pm 56$ & $2563$ & $2773$ & $1900_{-440}^{+430}$ \\ [2pt]
\hline
\end{tabular}
}
\end{table}

%----------------------------------------
\subsection{Emission spectrum of KELT-16 b}
\label{sec:emission_spectrum} %
\begin{figure}
\centering
$\begin{array}{cc}
\includegraphics[width=9cm]{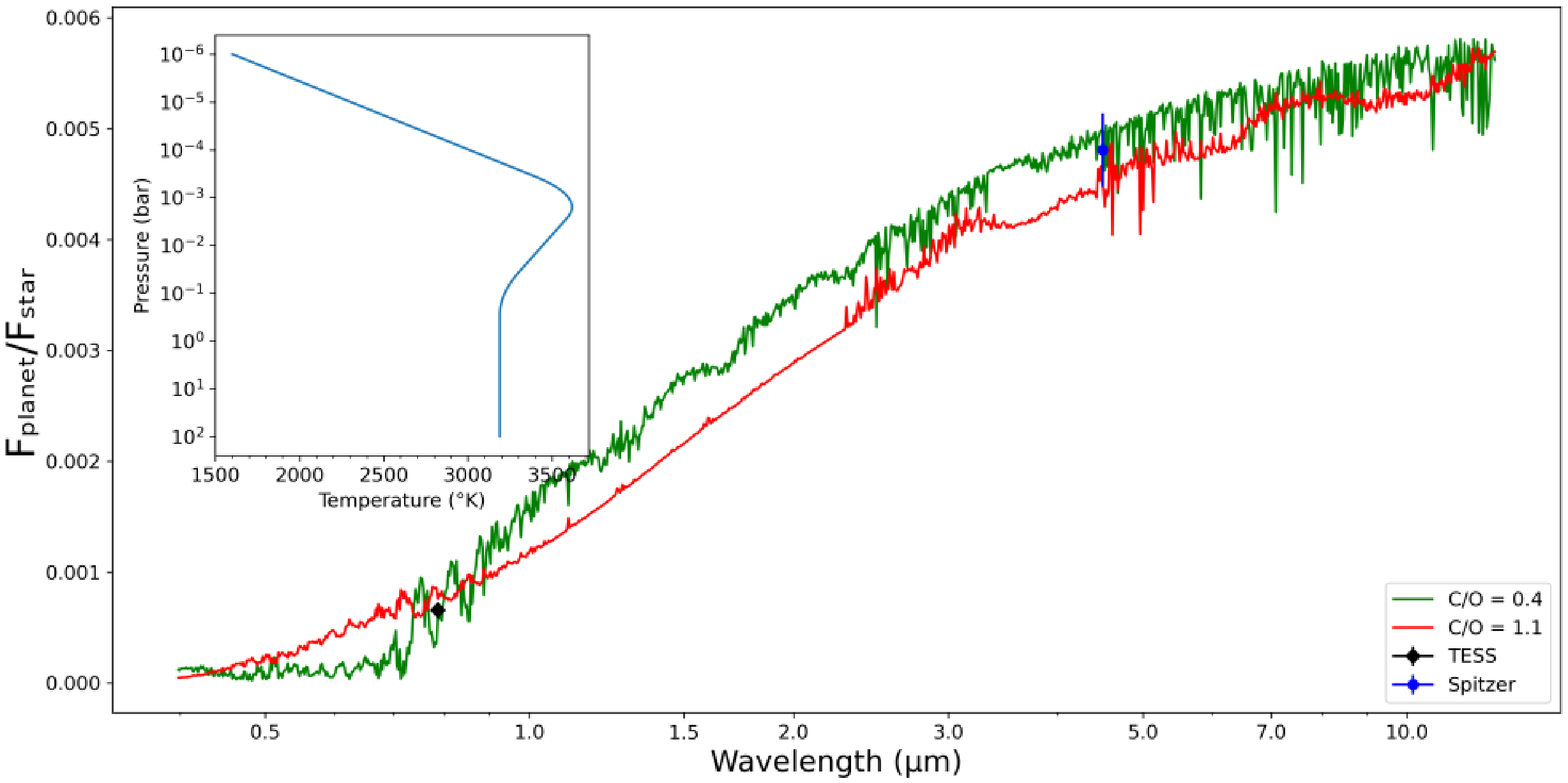} \\
\includegraphics[width=9cm]{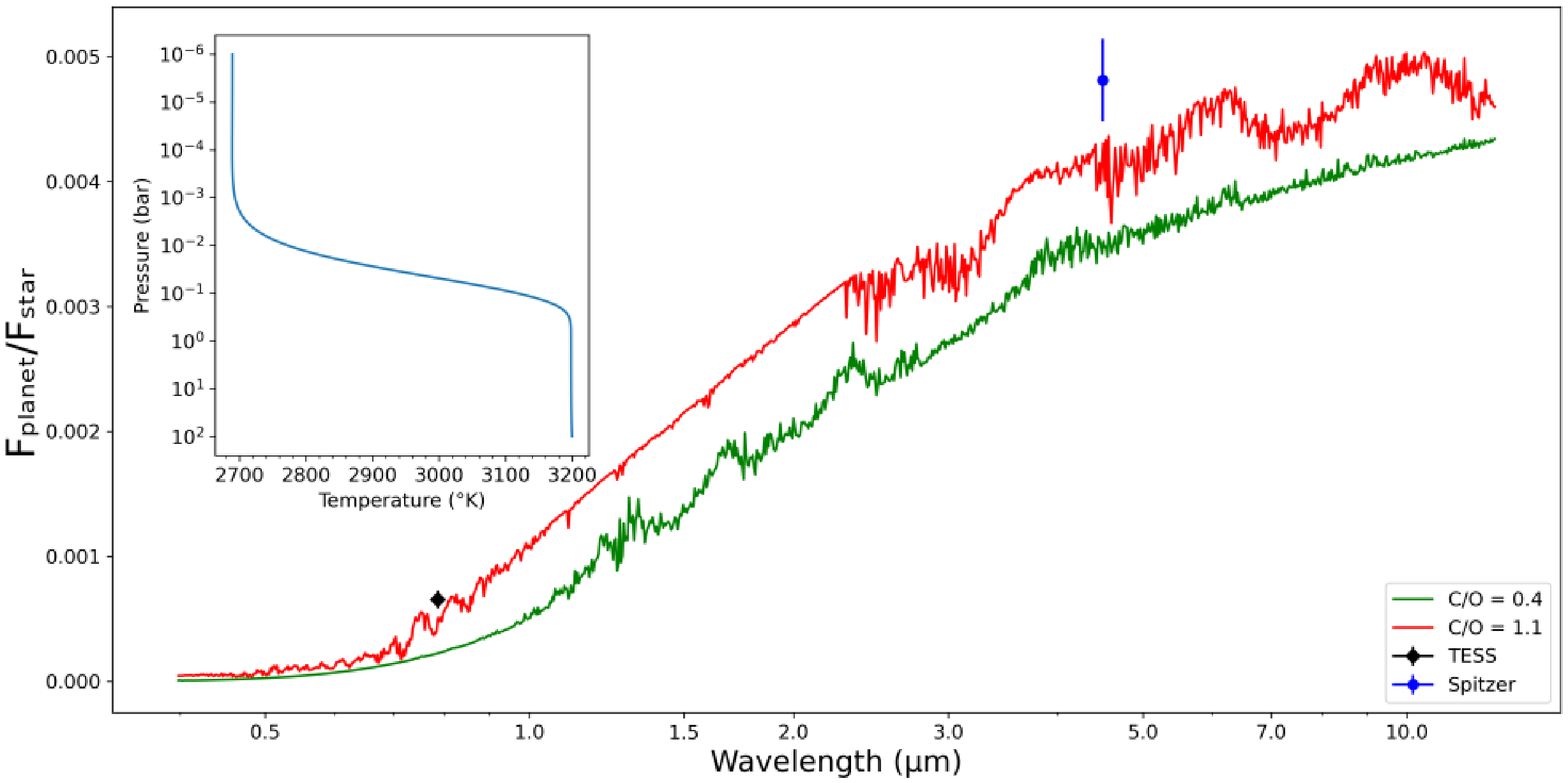} \\
\end{array}$
\caption{Dayside emission spectra of KELT-16b in terms of the planet-to-star flux ratio, with a specimen atmosphere rich in hydrogen and with two different C/O ratios and temperature profiles. The spectra in green and red correspond to ${\rm C/O}=0.4$ and ${\rm C/O}=1.1$ atmospheres, respectively, with a simple mixture of H$_{2}$O, CO, CH$_{4}$, CO$_{2}$, C$_{2}$H$_{2}$, HCN and TiO. The temperature profiles are shown in the upper left inner panels, the top one being a temperature inversion model with a peak temperature at 3600\,K, while the bottom one is the \citet{guillot:2010} model. The black diamond is the expected day flux ratio contribution from this paper with TESS data, while the blue dot is the calculated planet flux ratio with Spitzer from \citet{bell:2021}. The bandpass integrated model points are shown in the coloured squares. The two curves at the bottom of the panels show the TESS and the Spitzer $4.5\,\mu$m photometric bandpasses.}
\label{fig:emission}
\end{figure}

%----------------------------------------
In order to compare the calculated flux ratio at the TESS wavelength (centred on 786.5\,nm) with that at the Spitzer wavelength (centered on 4500\,nm) from \citet{bell:2021}, we used the Taurex python package \citep{alrefaie:2019} and generated four emission-spectrum models (Fig.~\ref{fig:emission}), which are based on the KELT-16 planetary-system parameters (Table~\ref{tab:finalparameters}), with two different atmosphere compositions\footnote{These atmosphere compositions are the same used by \citet{madhusudhan:2012} for WASP-19\,b, which has an equilibrium temperature and radius similar to those of KELT-16\,b.} (oxygen-dominated, ${\rm C/O}=0.4$ in green; carbon-dominated, ${\rm C/O}=1.1$ in red) and two different temperature profiles: a temperature inversion model for the top panel and the \citet{guillot:2010} model for the bottom panel. In every emission model, we set a constant quench temperature ($T = 3190$\,K), i.e. the temperature of the lower atmosphere ($P \sim 1$\,bar), as it fits very well the planet flux ratio calculated with TESS data. However, from an inspection of the two panels, only the inversion-temperature case manages to also fit the flux ratio from Spitzer with a peak temperature of $3600$\,K. 
%

%----------------------------------------
\section{Summary and conclusions}
\label{sec:conclusions} %
%----------------------------------------
In this work we studied the physical and orbital properties of the ultra-hot Jupiter KELT-16\,b, which is one of the few giant transiting exoplanets with $P_{\rm orb} < 1$\,day and belongs to a group of exoplanets that are the most favourable to detect orbital decay (see discussion in Sect.~\ref{sec:intro_2} and Sect.~\ref{sec:intro_3}). We reported the photometric monitoring of 28 transit events of KELT-16\,b, which were observed with five medium-class telescopes through four different optical passbands. Most of the transits were observed using the defocussing technique, achieving a photometric precision of 1\,mmag per observation in the best case (this unusual lack of good accuracy can be explained by the fact that there are no good comparison stars around KELT-16). Three transits were simultaneously observed with two different telescopes in different countries and one with three telescopes. In total, we collected 34 new light curves that were modelled with the {\sc jktebop} code  (see Figs.~\ref{fig:CA_lcs}, \ref{fig:Lo_lcs}, \ref{fig:T100_lcs}, \ref{fig:others_lcs}). We also considered the TESS data and analysed 18 complete transits recorded by this space telescope (see Fig.~\ref{fig:Fig_TESS_lc}). Our principal results are as follows.\\ 

$\bullet$ \ \ \ We estimated the mid-transit time for each of the transit event of KELT-16\,b that we presented. These timings were joined to others already published, obtaining a final list of 69 epochs (Table~\ref{tab:transit_times}), which we used for updating the ephemeris of the orbital period and the mid-transit time. We also searched for evidence of a decrease in its orbital period. Our analysis shows the possibility of TTVs in this planetary system, the values of the $\chi^{2}_{\nu}$, AIB and BIC of the orbital-decay ephemeris model being similar to the constant orbital-period model. Longer monitoring of KELT-16\,b transits are needed to obtain robust indications that the orbit of KELT-16\,b is decaying. Based on the current data and assuming that the period is not changing, we can set a limit of $Q^{\prime}_{\star}>(1.9 \pm 0.8) \times 10^5$ with $95\%$ confidence. \\

$\bullet$ \ \ \ We have used the TESS and the new ground-based light curves to refine the physical parameters of the KELT-16 planetary system. Our results are shown in Table~\ref{tab:finalparameters} and, in general, are in good agreement with those obtained by \citet{oberst:2017} but more precise. \\

$\bullet$ \ \ \ As stressed by \citet{oberst:2017}, its ultra-short period and the high irradiation make KELT-16\,b a benchmark target for atmospheric studies. Taking advantage of our multi-band photometric observations, we reconstructed a low-resolution optical transmission spectrum of the planet. We found a small variation of the planet's radius, which suggests the presence of strong absorbers in the optical, as expected, but at a low significance. More precise observations are mandatory to robustly confirm this indication. \\

$\bullet$ \ \ \ Using the TESS data, we reconstructed the phase curve of the KELT-16 adopting the periodicity and the time of transit shown in  Table~\ref{tab:finalparameters}. We simultaneously fitted the transit and occultation and estimated the flux ratio of the planet over its parent star and estimated the temperature of both the day- and night-side of the planet ($3190 \pm 61$\,K and $2668 \pm 56$\,K, respectively). Moreover, we found that KELT-16\,b has a phase offset of $ 25 \pm 14$\,$^{\circ}$E. These results (see Table~\ref{tab:posterior_values}), which are based on TESS data, are compatible with those found with Spitzer data \citep{bell:2021}. \\

$\bullet$ \ \ \ We compared the flux ratio at the TESS wavelength with that at the Spitzer wavelength and generated several emission-spectrum models for probing the chemical composition of the planet's atmosphere. We found that an atmosphere with temperature inversion is favoured, with a slight preference for an oxygen-dominated rather than a carbon-dominated composition. Again, many more measurements at different wavelength are needed to confirm this results.

\section*{Acknowledgements}
This paper is based on observations collected with ($i$) the Zeiss 1.23\,m telescope at the Centro Astron\'{o}mico Hispano Alem\'{a}n (CAHA) at Calar Alto, Spain; ($ii$) the Cassini 1.52\,m telescope, run by INAF--Osservatorio Astrofisico e Scienza dello Spazio di Bologna at Loiano, Italy; ($iii$) the T100 1.0\,m telescope at the TUG observatory, Turkey; ($iv$) the SPM 84\,cm telescope at the National Astronomical observatory, Mexico; ($v$) the 80\,cm telescope at the OARPAF observatory, Italy.
Operations at the Calar Alto telescopes are jointly performed by  Junta de Andaluc\'{i}a and the Instituto de Astrof\'{i}sica de Andaluc\'{i}a (CSIC) in Granada, Spain.
We thank Roberto Gualandi for his technical assistance at the Cassini telescope.
We thank the support astronomers of CAHA for their technical assistance at the Zeiss telescope.
GD acknowledges support from CONICYT project Basal AFB-170002. OB acknowledges the support by The Scientific and Technological Research Council of Turkey (T\"UB\.{I}TAK) with the project 118F042. We thank T\"UB\.{I}TAK for the partial support in using T100 telescope with the project numbers 19AT100-1471 and 18BT100-1334.
We thank the anonymous referee for their useful criticisms and suggestions that helped to improve the quality of this work.
The following internet-based resources were used in research for this paper: the ESO Digitized Sky Survey; the NASA Astrophysics Data System; the SIMBAD database operated at CDS, Strasbourg, France; the arXiv scientific paper preprint service operated by the Cornell University.

%%%%%%%%%%%%%%%%%%%%%%%%%%%%%%%%%%%%%%%%%%%%%%%%%%
\section*{Data Availability}
The reduced light curves presented in this work will be made available at the CDS (http://cdsweb.u-strasbg.fr/).


%\bibliographystyle{mnras}

%%%%%%%%%%%%%%%%%%%%%%%%%%%%%%%%%%%%%%%%%%%%%%%%%%

%%%%%%%%%%%%%%%%% APPENDICES %%%%%%%%%%%%%%%%%%%%%

\appendix
\section{Simultaneous light curves}
\label{sec:sim_light_curves}
\begin{figure}[H]
\includegraphics[width=\hsize]{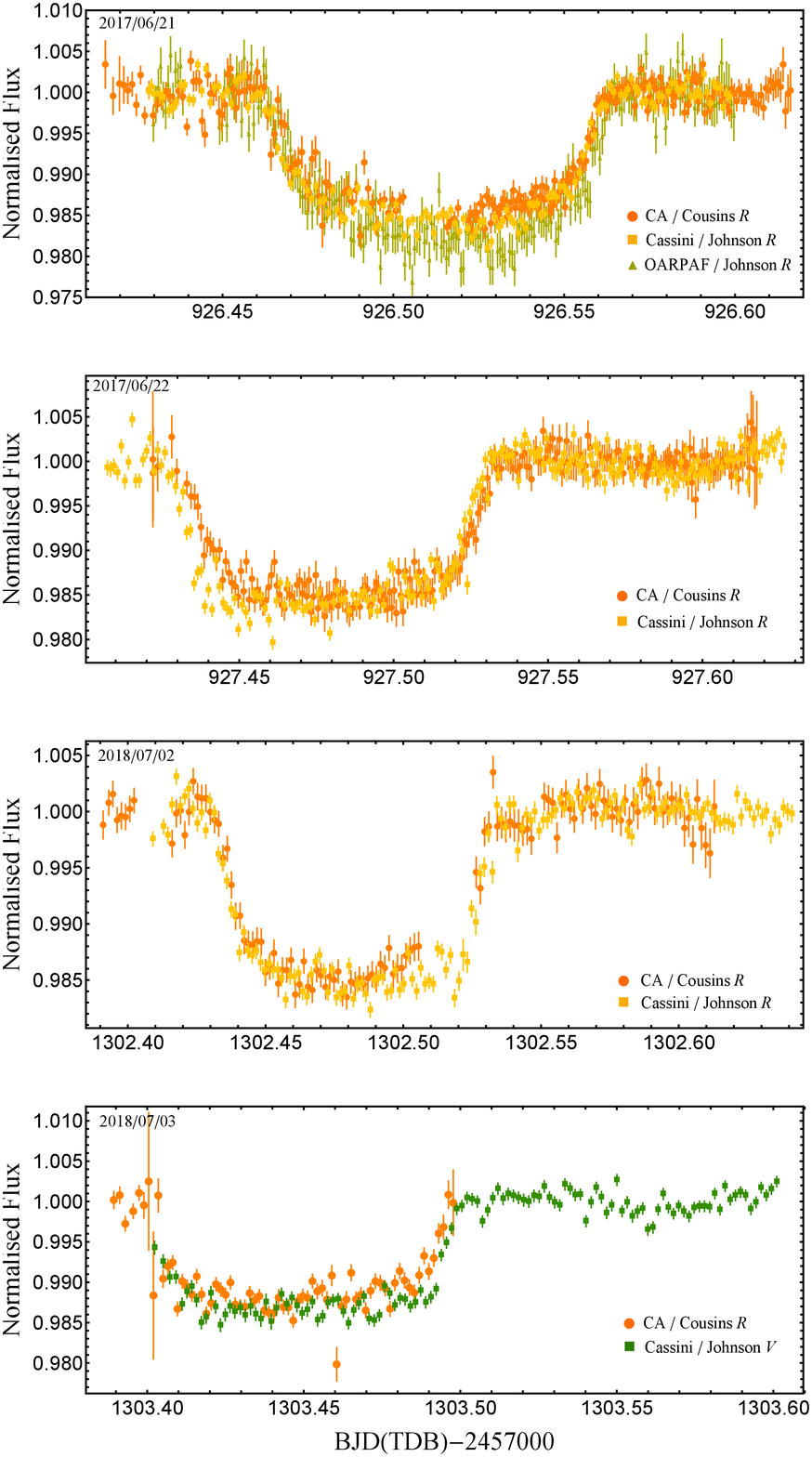}
\caption{The fours transit events that were simultaneously observed with three telescopes (top panel) and two telescopes bottom panels.}
\label{fig:simul_lcs}
\end{figure}
\section{Zoom in to the O--C plot}
\label{sec:zoom_o-c_plot}
\begin{figure}[H]
\includegraphics[width=\hsize]{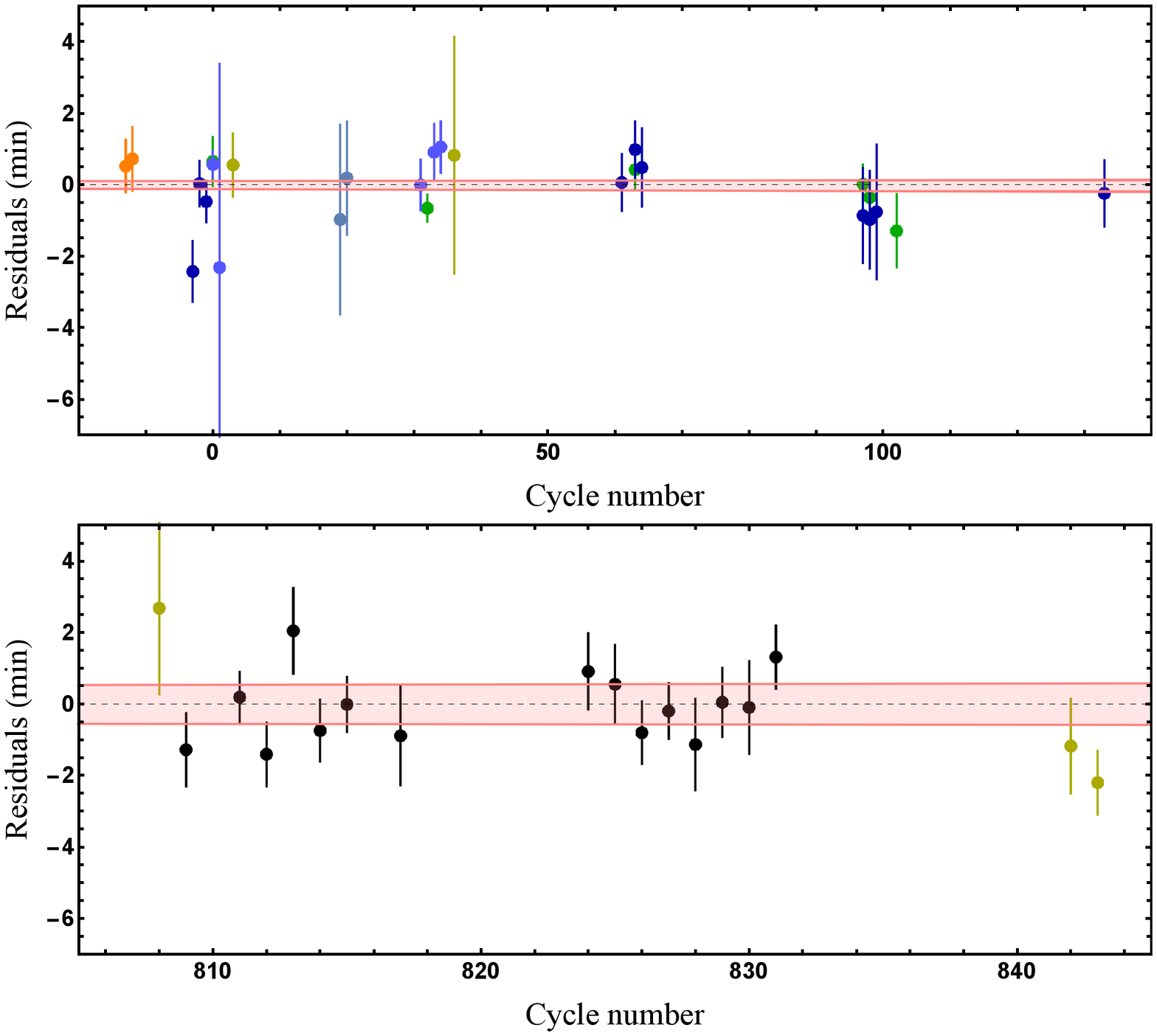}
\caption{Zooms in to the best sampled region (top panel) and in that covered by TESS (bottom panel). The points are coloured as in Fig~\ref{fig:Fig_O-C}.}
\label{fig:zoom}
\end{figure}
\section{Fit of the phase curve}
\label{sec:Distributions_of_the_posterior_values}
\begin{figure}
\includegraphics[width=\hsize]{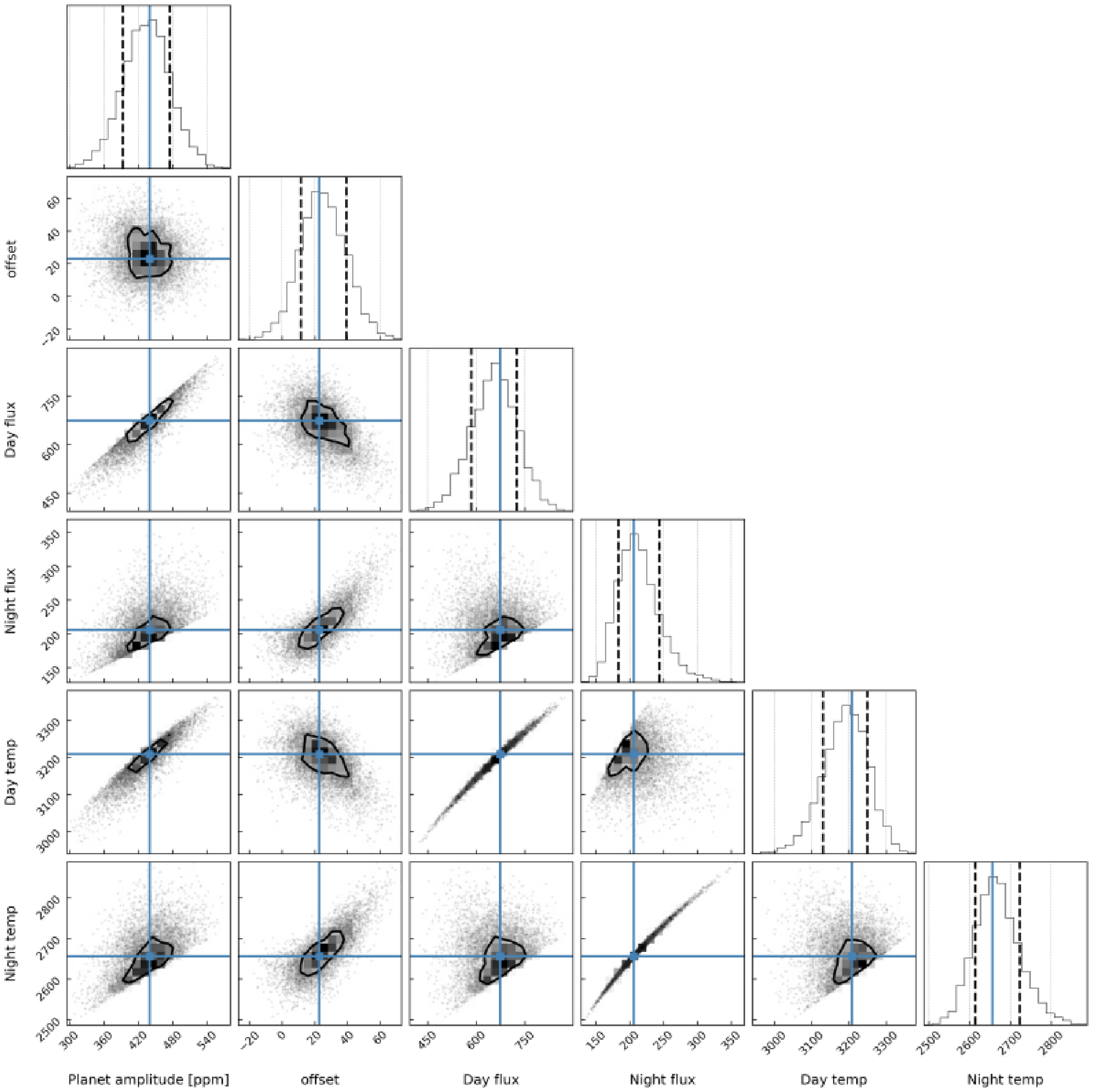}
\caption{Distributions of the posterior values over the model parameters from a Starry computation with 8000 draw iterations. Mean values are highlighted with blue lines, while the contour levels of the joint probability densities have been drawn at 1\,$\sigma$ distance.}
\label{fig:Fig_distribution}
\end{figure}

%%%%%%%%%%%%%%%%%%%%%%%%%%%%%%%%%%%%%%%%%%%%%%%%%%

% Don't change these lines
\bsp	% typesetting comment
\label{lastpage}
\end{document}